\documentclass[aps,prc,twocolumn,superscriptaddress]{revtex4-1}

\usepackage[utf8]{inputenc}
\usepackage[english]{babel}
\usepackage{bm}
\usepackage{graphicx}
\usepackage{epstopdf}
\usepackage{amsmath}
\usepackage{amssymb}
\usepackage{booktabs}
\usepackage{enumitem}
\usepackage{color}

\begin{document}
% Activate line numbers
%\linenumbers

\title{Measurements of the ion fraction and mobility of alpha and beta decay products in liquid xenon using EXO-200}

\newcommand{\Alabama}{\affiliation{Department of Physics and Astronomy, University of Alabama, Tuscaloosa, Alabama 35487, USA}}
\newcommand{\Alberta}{\affiliation{University of Alberta, Edmonton, Alberta, Canada}}
\newcommand{\Bern}{\affiliation{LHEP, Albert Einstein Center, University of Bern, Bern, Switzerland}}
\newcommand{\CALTECH}{\affiliation{Kellogg Lab, Caltech, Pasadena, California 91125, USA}}
\newcommand{\Carleton}{\affiliation{Physics Department, Carleton University, Ottawa, Ontario K1S 5B6, Canada}}
\newcommand{\CSU}{\affiliation{Physics Department, Colorado State University, Fort Collins, Colorado 80523, USA}}
\newcommand{\Drexel}{\affiliation{Department of Physics, Drexel University, Philadelphia, Pennsylvania 19104, USA}}
\newcommand{\Duke}{\affiliation{Department of Physics, Duke University, and Triangle Universities Nuclear Laboratory (TUNL), Durham, North Carolina 27708, USA}}
\newcommand{\IBS}{\affiliation{IBS Center for Underground Physics, Daejeon, Korea}}
\newcommand{\IHEP}{\affiliation{Institute of High Energy Physics, Beijing, China}}
\newcommand{\Illinois}{\affiliation{Physics Department, University of Illinois, Urbana-Champaign, Illinois 61801, USA}}
\newcommand{\Indiana}{\affiliation{Physics Department and CEEM, Indiana University, Bloomington, Indiana 47405, USA}}
\newcommand{\ITEP}{\affiliation{Institute for Theoretical and Experimental Physics, Moscow, Russia}}
\newcommand{\Laurentian}{\affiliation{Department of Physics, Laurentian University, Sudbury, Ontario P3E 2C6, Canada}}
\newcommand{\Maryland}{\affiliation{Physics Department, University of Maryland, College Park, Maryland 20742, USA}}
\newcommand{\Munich}{\affiliation{Technische Universit\"at M\"unchen, Physikdepartment and Excellence Cluster Universe, Garching 85748, Germany}}
\newcommand{\SDakota}{\affiliation{Physics Department, University of South Dakota, Vermillion, South Dakota 57069, USA}}
\newcommand{\Seoul}{\affiliation{Department of Physics, University of Seoul, Seoul, Korea}}
\newcommand{\SLAC}{\affiliation{SLAC National Accelerator Laboratory, Menlo Park, California 94025, USA}}
\newcommand{\Stanford}{\affiliation{Physics Department, Stanford University, Stanford, California 94305, USA}}
\newcommand{\Stony}{\affiliation{Department of Physics and Astronomy, Stony Brook University, SUNY, Stony Brook, New York 11794, USA}}
\newcommand{\TRIUMF}{\affiliation{TRIUMF, Vancouver, BC, Canada}}
\newcommand{\UMass}{\affiliation{Amherst Center for Fundamental Interactions and Physics Department, University of Massachusetts, Amherst, MA 01003, USA}}
\newcommand{\WIPP}{\affiliation{Waste Isolation Pilot Plant, Carlsbad, New Mexico 88220, USA}}
\author{J.B.~Albert}\Indiana
\author{D.J.~Auty}\altaffiliation{Now at University of Alberta, Edmonton, Alberta, Canada}\Alabama
\author{P.S.~Barbeau}\Duke
\author{D.~Beck}\Illinois
\author{V.~Belov}\ITEP
\author{M.~Breidenbach}\SLAC
\author{T.~Brunner}\Stanford
\author{A.~Burenkov}\ITEP
\author{G.F.~Cao}\IHEP
\author{C.~Chambers}\CSU
\author{B.~Cleveland}\altaffiliation{Also SNOLAB, Sudbury ON, Canada}\Laurentian
\author{M.~Coon}\Illinois
\author{A.~Craycraft}\CSU
\author{T.~Daniels}\SLAC
\author{M.~Danilov}\ITEP
\author{S.J.~Daugherty}\Indiana
\author{C.G.~Davis}\altaffiliation{Now at the Naval Research Lab, Washington D.C., USA}\Maryland
\author{J.~Davis}\SLAC
\author{S.~Delaquis}\Bern
\author{A.~Der Mesrobian-Kabakian}\Laurentian
\author{R.~DeVoe}\Stanford
\author{T.~Didberidze}\Alabama
\author{A.~Dolgolenko}\ITEP
\author{M.J.~Dolinski}\Drexel
\author{M.~Dunford}\Carleton
\author{W.~Fairbank Jr.}\CSU
\author{J.~Farine}\Laurentian
\author{W.~Feldmeier}\Munich
\author{P.~Fierlinger}\Munich
\author{D.~Fudenberg}\Stanford
\author{R.~Gornea}\Bern
\author{K.~Graham}\Carleton
\author{G.~Gratta}\Stanford
\author{C.~Hall}\Maryland
\author{M.~Hughes}\Alabama
\author{M.J.~Jewell}\Stanford
\author{X.S.~Jiang}\IHEP
\author{A.~Johnson}\SLAC
\author{T.N.~Johnson}\Indiana
\author{S.~Johnston}\UMass
\author{A.~Karelin}\ITEP
\author{L.J.~Kaufman}\Indiana
\author{R.~Killick}\Carleton
\author{T.~Koffas}\Carleton
\author{S.~Kravitz}\Stanford
\author{A.~Kuchenkov}\ITEP
\author{K.S.~Kumar}\Stony
\author{D.S.~Leonard}\IBS
\author{C.~Licciardi}\Carleton
\author{Y.H.~Lin}\Drexel
\author{J.~Ling}\Illinois
\author{R.~MacLellan}\SDakota
\author{M.G.~Marino}\Munich
\author{B.~Mong}\altaffiliation{Corresponding author: \href{mailto:brian.e.mong@gmail.com}{brian.e.mong@gmail.com}}\Laurentian
\author{D.~Moore}\Stanford
\author{R.~Nelson}\WIPP
\author{K.~O'Sullivan}\altaffiliation{Now at Yale University, New Haven, CT, USA}\Stanford
\author{A.~Odian}\SLAC
\author{I.~Ostrovskiy}\Stanford
\author{A.~Piepke}\Alabama
\author{A.~Pocar}\UMass
\author{C.Y.~Prescott}\SLAC
\author{A.~Robinson}\Laurentian
\author{P.C.~Rowson}\SLAC
\author{J.J.~Russell}\SLAC
\author{A.~Schubert}\Stanford
\author{D.~Sinclair}\TRIUMF\Carleton
\author{E.~Smith}\Drexel
\author{V.~Stekhanov}\ITEP
\author{M.~Tarka}\Stony
\author{T.~Tolba}\Bern
\author{R.~Tsang}\Alabama
\author{K.~Twelker}\Stanford
\author{J.-L.~Vuilleumier}\Bern
\author{A.~Waite}\SLAC
\author{J.~Walton}\Illinois
\author{T.~Walton}\CSU
\author{M.~Weber}\Stanford
\author{L.J.~Wen}\IHEP
\author{U.~Wichoski}\Laurentian
\author{J.D.~Wright}\UMass 
\author{J.~Wood}\WIPP
\author{L.~Yang}\Illinois
\author{Y.-R.~Yen}\Drexel
\author{O.Ya.~Zeldovich}\ITEP

\collaboration{EXO-200 Collaboration}

\date{\today}  

\begin{abstract}
%Targets: NIMA, PRC, JPhysD, PRX
Alpha decays in the EXO-200 detector are used to measure the fraction of charged $^{218}\mathrm{Po}$ and $^{214}\mathrm{Bi}$ daughters created from alpha and beta decays, respectively. % in liquid xenon (LXe).
$^{222}\mathrm{Rn}$ alpha decays in liquid xenon (LXe) are found to produce $^{218}\mathrm{Po}^{+}$ ions $50.3 \pm 3.0\%$ of the time, while the remainder of the $^{218}\mathrm{Po}$ atoms are neutral. 
The fraction of $^{214}\mathrm{Bi}^{+}$ from $^{214}\mathrm{Pb}$ beta decays in LXe is found to be $76.4 \pm 5.7\%$, inferred from the relative rates of $^{218}\mathrm{Po}$ and $^{214}\mathrm{Po}$ alpha decays in the LXe.
The average velocity of $^{218}\mathrm{Po}$ ions is observed to decrease for longer drift times. %, which is explained by having two different mobilities.
Initially the ions have a mobility of $0.390 \pm 0.006~\mathrm{cm}^2/(\mathrm{kV}~\mathrm{s})$, and at long drift times the mobility is $0.219 \pm 0.004~\mathrm{cm}^2/(\mathrm{kV}~\mathrm{s})$.
Time constants associated with the change in mobility during drift of the $^{218}\mathrm{Po}^{+}$ ions are found to be proportional to the electron lifetime in the LXe.
%, with the latter being negligible at nominal operating conditions.
\end{abstract}

%\keywords{}

\maketitle

%\listoffigures
%\listoftables
%\tableofcontents

\section{Introduction}\label{sec:Introduction} 

EXO-200 discovered the two-neutrino double beta decay in $^{136}\mathrm{Xe}$ \cite{Ackerman:2011gz}, and subsequently made a precision measurement of the half-life of the decay \cite{Albert2014_2vbb}.
The experiment is currently searching for neutrinoless double beta decay ($0 \nu \beta \beta$), and has produced limits on the half-life of this process in excess of $10^{25}~\mathrm{years}$ \cite{Auger:2012ar, Albert2014_0vbb} using $110~\mathrm{kg}$ of liquid Xe (LXe) enriched to $80.6\%$ in the isotope $^{136}\mathrm{Xe}$ \cite{Auger:2012gs}.
Analyses of alpha decays were used to establish the contributions of internal backgrounds (e.g. from the uranium series) to the $0 \nu \beta \beta$ measurement \cite{exoBG2015}.
In this paper, these alpha decays are used to make novel measurements of ion fractions resulting from alpha and beta decay and to determine the $^{218}\mathrm{Po}$ ion mobility in LXe. 
These measurements may also be of interest to other experiments that utilize a noble liquid TPC, such as dark matter searches and long-baseline neutrino oscillation experiments.
Radon and its by-products are also found in these detectors.

\begin{figure}
\includegraphics[width=0.48\textwidth]{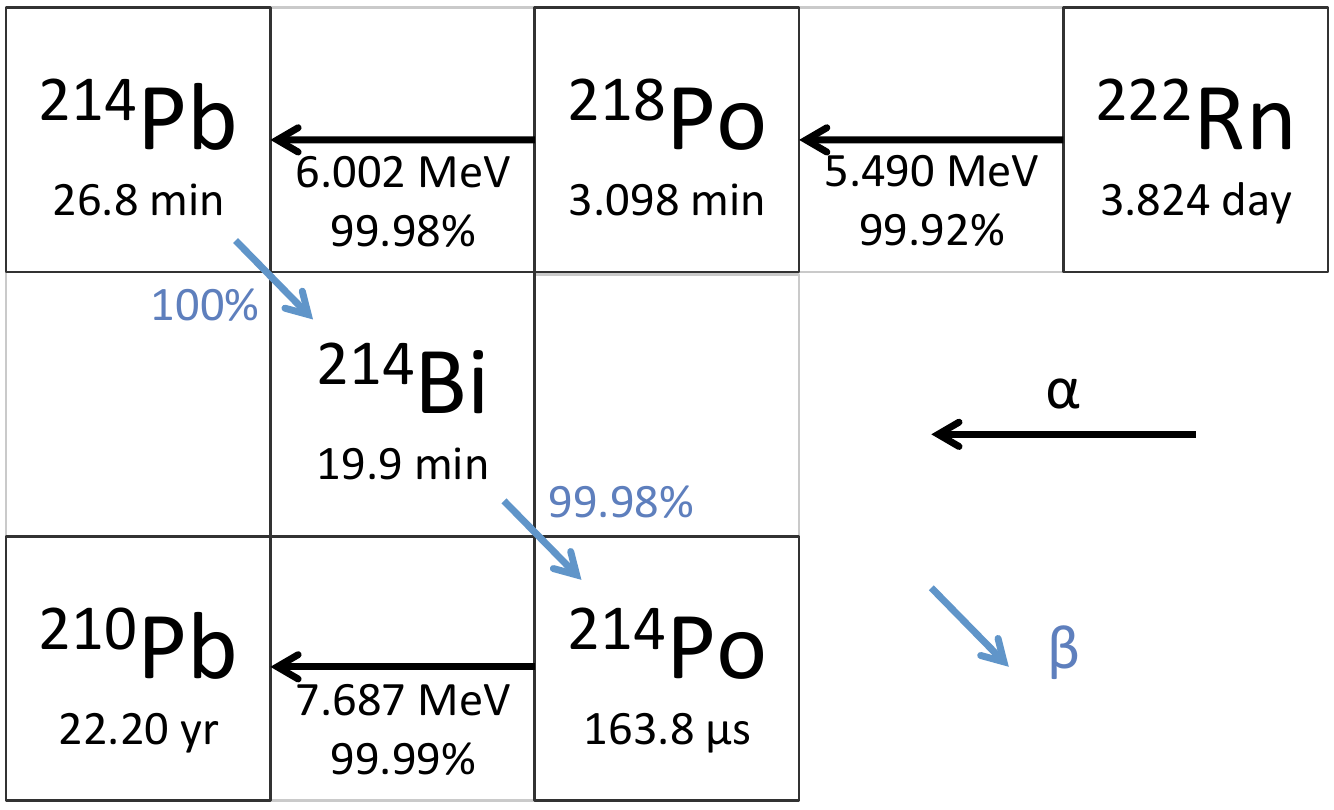}
\caption{(Color online) Partial diagram of the uranium series. 
The alpha particle energies are shown along with their corresponding relative intensities. 
The beta decays are shown with their branching ratios.}
\label{fig:Rn222_decay_chain}
\end{figure}

The relevant portion of the uranium series is shown in Fig.~\ref{fig:Rn222_decay_chain}.
This part of the series contains three alpha and two beta decays that occur within a relatively short time period.
% and therefore can be 
%are candidates for coincidence studies.
%In this paper, measurements of the $^{222}\mathrm{Rn}$ content in the liquid xenon are presented.
Using the EXO-200 detector it is possible to match many $^{222}\mathrm{Rn}$ decays to their daughter $^{218}\mathrm{Po}$ decays.
These matched pairs are then used to measure the ion fraction and ion mobility of $^{218}\mathrm{Po}$ resulting from the alpha decay.
Finally the activities of the three alpha decays,$^{222}\mathrm{Rn}$, $^{218}\mathrm{Po}$, and $^{214}\mathrm{Po}$, are used to determine the ion fraction of $^{214}\mathrm{Bi}$ produced from beta decay.

The fraction of ions created from beta decays and the behavior of ions in the detector is of particular interest for nEXO, the next-generation EXO experiment, which is expected to contain 5 tons of LXe enriched in $^{136}\mathrm{Xe}$.
A possible second phase of nEXO may include a system to ``tag'' the $^{136}\mathrm{Ba}$ daughter of the $0 \nu \beta \beta$ decay in order to eliminate all background except a negligible contribution from the two-neutrino mode \cite{Twelker2014,Mong2015}.
%Whether the $^{136}\mathrm{Ba}$ daughter is mainly an ion or a neutral atom, what the charge of the ion is in LXe, and how the Ba will move in the detector over the time scales required to detect it are important unanswered questions relevant to the design and implementation of barium tagging.
There are several important unanswered questions relevant to the implementation of Ba tagging: 1) whether the $^{136}\mathrm{Ba}$ daughter is mainly an ion or a neutral atom, 2) what the charge of the ion is in LXe, and 3) how the Ba will move in the detector over time scales required to capture it.
One way to probe the behavior of ions and atoms in the detector is to track events as they decay through the uranium series.

To our knowledge there are no previous determinations of the daughter ion fractions of alpha or beta decay within liquid noble gases.
In related work with superfluid helium, the ion fraction of $^{219}\mathrm{Rn}$ recoils into the liquid from $^{223}\mathrm{Ra}$ alpha decay occurring on a surface has been measured to be $5.36 \pm 0.13\%$ and $1.04 \pm 0.06\%$ for two different surface geometries \cite{purushothaman2008}, and the surviving ion fractions for high energy external beams of $^{12}\mathrm{B}$, $^{12}\mathrm{N}$ and $^{8}\mathrm{Li}$ have been reported as $30\%$, $10\%$, and $35\%$, respectively \cite{takahashi2003}.
The $^{218}\mathrm{Po}$ ion fraction from $^{222}\mathrm{Rn}$ decays in helium and argon gas has been measured to be $59\pm4\%$ and $48\pm5\%$ respectively at atmospheric pressure \cite{Howard1991}.
The ion fraction of $^{218}\mathrm{Po}$ has also been measured in air at atmospheric pressure to be $87.3 \pm 1.6\%$ \cite{Pagelkopf2003}, and $88\%$ \cite{Hopke1996}.

Mobility measurements have been reported for four alkaline earth ions in LXe including $\mathrm{Ba}^+$ \cite{Jeng2009}, as well as $\mathrm{Tl}^+$  \cite{Jeng2009,Walters2003} and $\mathrm{Th}^+$ \cite{Wamba2005}.
These mobilities range from $0.133~\mathrm{cm}^2/(\mathrm{kV}~\mathrm{s})$ to $0.280~\mathrm{cm}^2/(\mathrm{kV}~\mathrm{s})$.

%The measurements in this paper may be of interest to other experiments that utilize a noble liquid TPC, such as dark matter searches and long-baseline neutrino oscillation experiments.
%Radon and its by-products are also found in these detectors. 

%%%%%%%%%%%%%%%%%%%%%%%%%%%%%%%%%%%%%%%%%%%%%%%%%%%%%%%%%%%%%%%%%%%%%%%%%%%%%%%%%%%%%%%
\section{EXO-200 detector}\label{sec:Detector}
\begin{figure}
\begin{center}
\includegraphics[width=0.48\textwidth]{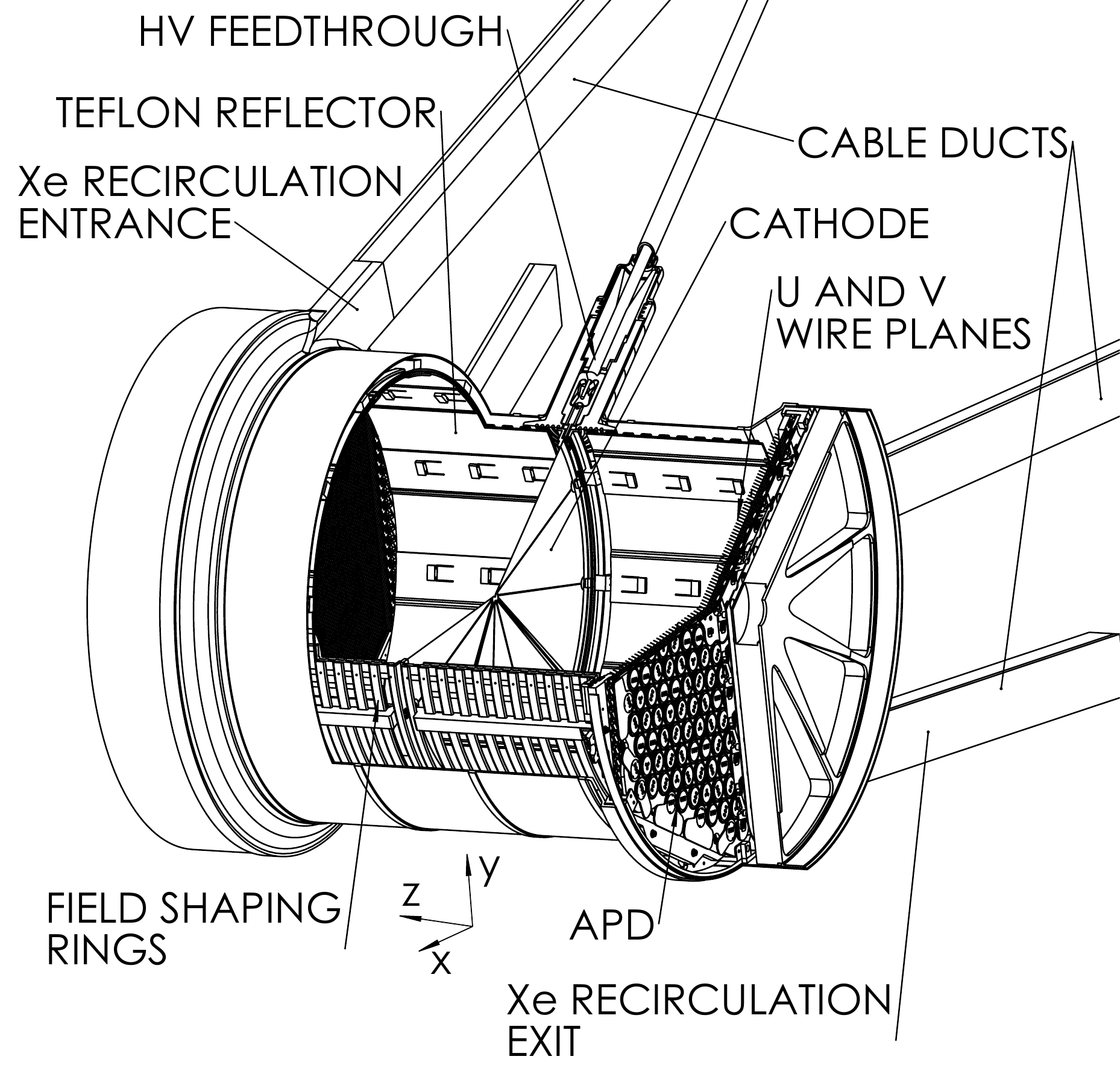}
\end{center}
\caption{Cutaway view of the EXO-200 detector highlighting several important components of the TPC.}
\label{fig:TPC_diagram}
\end{figure}

%%%%%%%%%%%%%%%%%%%%%%%%%%%%%%%%%%%%%%%%%%%%%%%%%%%%%%%%%%%%%%%%%%%%%%%%%%%%%%%%%%%%%%%
EXO-200 is described in detail elsewhere~\cite{Auger:2012gs}. 
To summarize, the detector (shown in Fig.~\ref{fig:TPC_diagram}) consists of two symmetric back-to-back cylindrical time projection chambers (TPC) with a common mesh cathode.
The TPCs are labeled $\mathrm{TPC1}$ and $\mathrm{TPC2}$ for the $+z$ and $-z$ sides of the cathode respectively, (where $z$ is along the axis of symmetry).
The detector is filled with LXe that is isotopically enriched to $80.6\%$ in $^{136}\mathrm{Xe}$, with the remainder being primarily $^{134}\mathrm{Xe}$.
The density of this LXe is $3.03~\mathrm{g/cm^3}$.
LXe serves as both the double beta decay source and the detection medium, converting energy deposits into ionization and scintillation.
The LXe temperature in the detector is $166.6 \pm 0.2~\mathrm{K}$, and nominal pressure is $146\pm 6~\mathrm{kPa}$. %($1100\pm 40~\mathrm{torr}$).

The central cathode is set to a potential of $-8.0~\mathrm{kV}$.
At each anode two crossed wire planes are used to measure the drifted ionization. %, which are crossed at $60^{\circ}$.
The potential of each plane is set such that the charge drifts past the first wire plane (V-wires = $-780~\mathrm{V}$), inducing a signal, and is collected on the second wire plane (U-wires = $0~\mathrm{V}$).
These ionization signals are used to determine the event's $x$ and $y$ position(s). 
There are 38 U- and 38 V-wire channels in each TPC, each channel consisting of three adjacent wires connected together.
The wire-to-wire spacing in both wire planes is $0.3~\mathrm{cm}$. 
The total drift distance from the cathode to the V-wire plane is $19.2~\mathrm{cm}$. 
There is a $0.6~\mathrm{cm}$ gap between the V- and U-wire planes.
Behind the U-wires, by $0.6~\mathrm{cm}$, is a plane of large area avalanche photo diodes (APDs) which detect the scintillation light \cite{Neilson:2009kf}.
These APD planes are biased with $-1400~\mathrm{V}$.
A 3D simulation of the TPC has determined the drift field in the analysis region to be $380~\mathrm{V/cm}$, with some position dependence ($\pm 5~\mathrm{V/cm}$). 

The data acquisition system (DAQ) continually samples signals from each U, V and APD channel at $1~\mathrm{MHz}$.  
When a U-wire, APD, or APD-sum goes above appropriate thresholds, the DAQ records a $2048~\mathrm{\mu s}$ long waveform, with 1024 pre-trigger samples.
Later, during data processing, all waveforms are scanned for signals.
APD and wire signals are then grouped together using spatial and temporal information.
The time difference between scintillation and ionization signals is converted to the $z$ position of the event, using the electron drift speed of $1.71~\mathrm{mm/\mu s}$ \cite{Albert2014_2vbb}.
If two events occur within the maximum electron drift time ($116~\mathrm{\mu s}$), a clustering algorithm associates the charge signals to the scintillation signals that occurred closest in time.
Further information about data handling and reconstruction of events in EXO-200 can be found in \cite{Albert2014_2vbb}.

High purity conditions are maintained in the detector to maximize collection of charge and light.
This purity is achieved by continually recirculating xenon gas in a purification system consisting of a heater, a pump \cite{LePort:2011hy}, getter purifiers, and a condenser. 
The xenon flow is in the $-z$ direction with respect to the detector coordinates, with the input near the TPC1 anode, and output near the TPC2 anode. 
The purity of the LXe in each TPC is monitored using a calibration source deployed every few days. 
Purity is measured in terms of the electron lifetime, as determined from the attenuation of the charge signals produced by the calibration source as a function of drift distance.
The electron lifetime is typically maintained between $3$ and $5~\mathrm{ms}$.
A xenon feed and bleed system is used to fill and empty the detector, as well as to make adjustments to the xenon system pressure.
Occasional pauses in recirculation, for various operational reasons, cause temporary drops in the LXe purity. 
%Often such recirculation pauses coincide with a feed event, in which Xe is introduced to the system from storage, where it has not had the benefit of constant purification and therefore also reduce the electron lifetime.
These recirculation pauses often coincide with feed events, in which stored Xe is introduced into the system, which further reduces the electron lifetime since this Xe has not had the benefit of constant purification.
Feed events also introduce $^{222}\mathrm{Rn}$ into the detector from the gas storage system.
The source of this $^{222}\mathrm{Rn}$ is unknown, but is likely due to emanating component(s) or from known small leaks to atmosphere (or both) in the gas storage system.
The xenon gas purity in EXO-200 has been measured for some contaminants in the recirculation system (gas phase) using cold trap mass spectrometry during operation \cite{dobi2012xenon}, however these measurements are not necessarily representative of the contaminants in the liquid phase, nor are they exhaustive.

The experiment is installed underground at the Waste Isolation Pilot Plant near Carlsbad, NM.
The site has an overburden of $1585$ meters water equivalent which partially shields the detector from cosmic rays.
An active muon veto system is installed around the detector to help reject cosmogenic backgrounds.

%%%%%%%%%%%%%%%%%%%%%%%%%%%%%%%%%%%%%%%%%%%%%%%%%%%%%%%%%%%%%%%%%%%%%%%%%%%%%%%%%%%%%%%
\section{Analysis}
%%%%%%%%%%%%%%%%%%%%%%%%%%%%%%%%%%%%%%%%%%%%%%%%%%%%%%%%%%%%%%%%%%%%%%%%%%%%%%%%%%%%%%%
This analysis was performed using the data set from the latest $0\nu\beta\beta$ result \cite{Albert2014_0vbb} with the addition of runs near xenon feed events. 
These additional runs were not used in \cite{Albert2014_0vbb} due to the rapidly changing electron lifetime.
Nevertheless, they are useful here because they have a higher $^{222}\mathrm{Rn}$ content, and this analysis, relying primarily on scintillation signals, does not require high accuracy for charge energy.
Time periods following muons detected by the veto system or the TPC are not removed since cosmogenic backgrounds are unlikely to be confused as alpha decays.
%Dead times normally applied when muons are detected by the veto system or the TPC are not applied in this analysis because cosmogenic backgrounds are unlikely to be confused as alpha decays.
However, events containing muon tracks and events identified as noise are removed from the data set.  
The total live-time used for this analysis is $572.8~\mathrm{days}$. %\footnote{Computed from the DAQ, elapsed time.}.

Several corrections are applied to the data including wire-channel gain, electron lifetime, and V-wire shielding inefficiency (see \cite{Albert2014_2vbb} for more details).
For events that are fully reconstructed (have a U-, V-wire, and a scintillation signal), the APD gain variations and the position dependence of the scintillation light collection are corrected using a light map derived from calibration source data.
Events missing V-wire signals cannot be fully reconstructed in $x$-$y$ position, and therefore cannot be light map corrected.
All of the events used in this analysis have at least a scintillation and U-wire (charge collection) signal, which is sufficient to determine the $z$ position of the event. 

\subsection{Identifying alpha decays}
\begin{figure}
\includegraphics[width=0.48\textwidth]{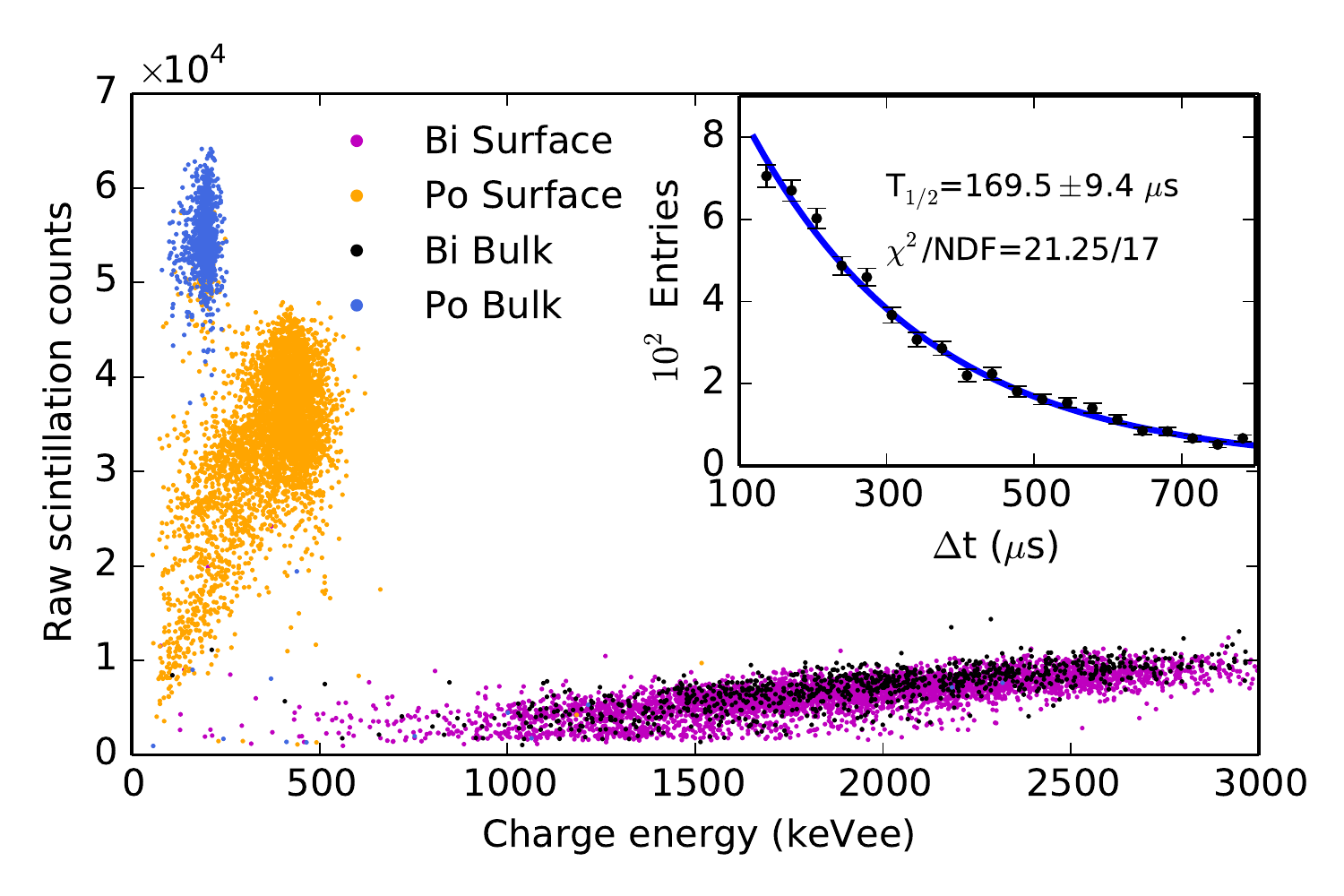}
\caption{(Color online) Raw scintillation versus electron equivalent charge energy ($\mathrm{keVee}$) of Bi-Po events. 
Different colors distinguish between  Bi and Po on surfaces and in the bulk LXe. 
Surface events are defined to be within $10~\mathrm{mm}$ of the cathode or anodes. 
The inset shows a histogram of the time delay between Bi-Po pairs.}
\label{fig:bipo_light_charge}
\end{figure}

Alphas interacting in the LXe are readily distinguished from beta and gamma interactions due to the large amount of scintillation relative to ionization. 
This difference is caused by the higher ionization density for tracks created by alpha particles, resulting in more recombination. 
This is demonstrated in Fig.~\ref{fig:bipo_light_charge} where ${^{214}\mathrm{Bi}} \rightarrow {^{214}\mathrm{Po}}$ (Bi-Po) coincidence events are plotted as scintillation counts versus charge energy.
The $^{214}\mathrm{Bi}$ decay (emitting a beta and several gammas) and $^{214}\mathrm{Po}$ (alpha decay) are easy to identify since the events occur in rapid succession ($\mathrm{T}_{1/2}=163.8~\mu \mathrm{s}$), usually in the same DAQ frame. 
Here ``bulk'' coincidences (blue and black) are identified by a $^{214}\mathrm{Po}$ alpha decay with a $z$ position between $10~\mathrm{mm}$ and $182~\mathrm{mm}$ from the cathode, while ``surface'' coincidences (magenta and green) are within $10~\mathrm{mm}$ of either the cathode or V-wire surfaces. 
Alpha decays on or near the cathode or anodes are observed to have a higher charge energy, and a correspondingly lower scintillation energy, than those which decay in the bulk xenon due to the higher local electric field.

\begin{figure}
\includegraphics[width=0.48\textwidth]{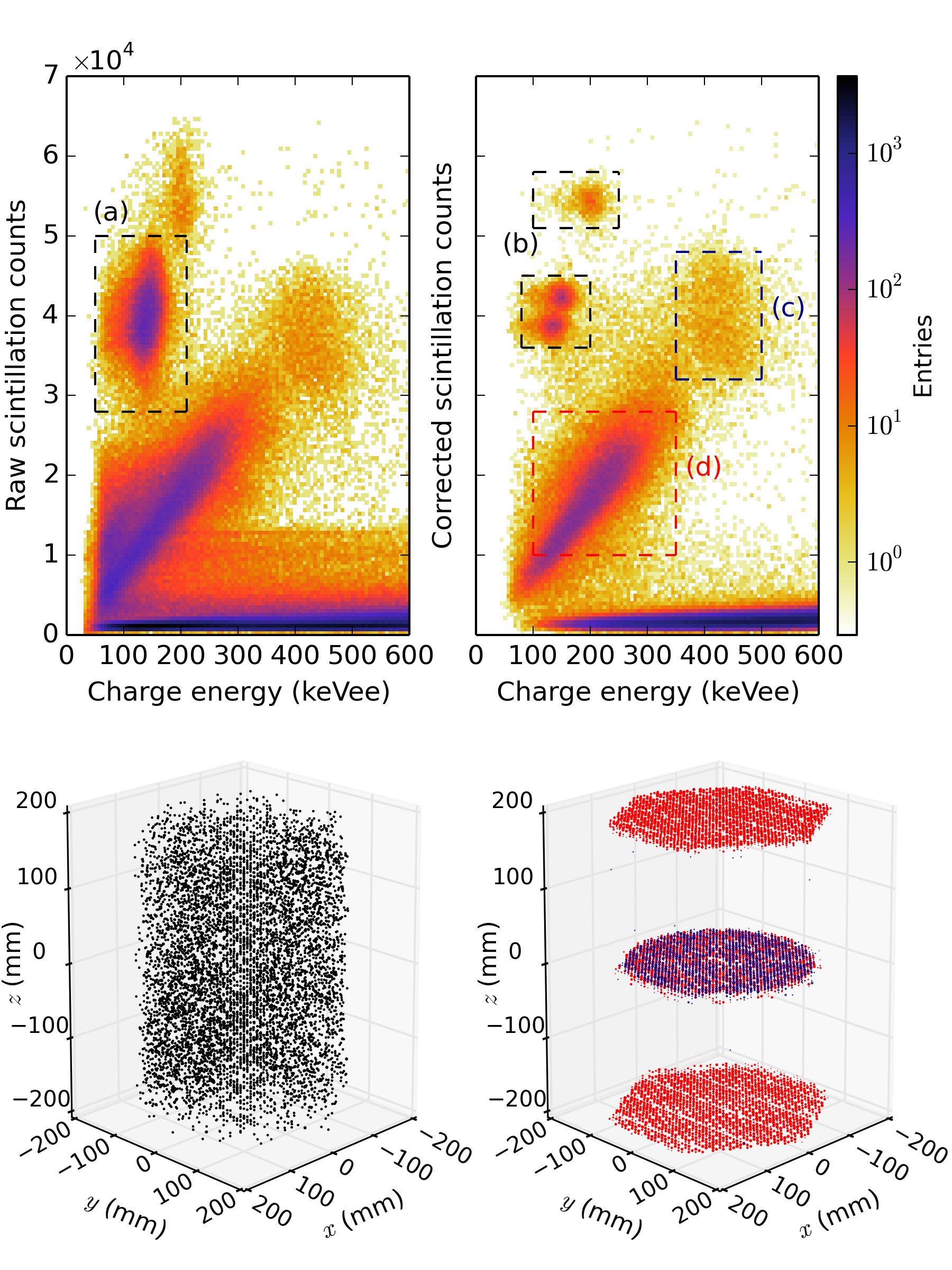}
\caption{(Color online) Upper-Left: raw scintillation counts versus charge energy histogram. The highlighted region (a) is used to monitor activity of the $^{222}\mathrm{Rn}$ chain in the LXe.
Upper-Right: fully reconstructed and light map corrected data. %with regions decays in the LXe (B) and on or near surfaces (C).
Lower-Left: positions of events in region (b). 
Lower-Right: positions of events in regions (c) (dark blue) and (d) (red).}
\label{fig:recon_light_charge_alphas}
\end{figure}

A scintillation versus charge histogram of the whole data set, highlighting the regions of interest with alpha events, is shown in Fig.~\ref{fig:recon_light_charge_alphas}, for both raw and light map corrected scintillation counts.
In the fully reconstructed and light map corrected data (upper right), the position of alpha events in the bulk (region B) and near surfaces (region C) is clearly separated, which is shown in the lower plots.
%The three alpha decays in the uranium series are identified with a mean scintillation of $3.88\times 10^4$, $4.24\times 10^4$, and $5.43\times 10^4$ counts for the $^{222}\mathrm{Rn}$, $^{218}\mathrm{Po}$, and $^{214}\mathrm{Po}$ decays respectively.
The three alpha decays in the uranium series are identified with as three separate peaks in Fig.~\ref{fig:recon_light_charge_alphas} region B, with mean scintillation of $38766$, $42377$, and $54254$ counts for the $^{222}\mathrm{Rn}$, $^{218}\mathrm{Po}$, and $^{214}\mathrm{Po}$ decays respectively.
These means are found to be linear to within $<0.03\%$ with respect to the alpha energies.
There are significantly fewer alpha events in the light map corrected data, due to missing V-wire signals.
%To maintain full statistics this analysis has been performed without requiring the V-wire signals by using the uncorrected scintillation and U-wire signals. 
To maintain full statistics, these events are analyzed using the uncorrected scintillation and U-wire signals, removing the requirement for a V-wire signal.

\begin{figure}
\includegraphics[width=0.48\textwidth]{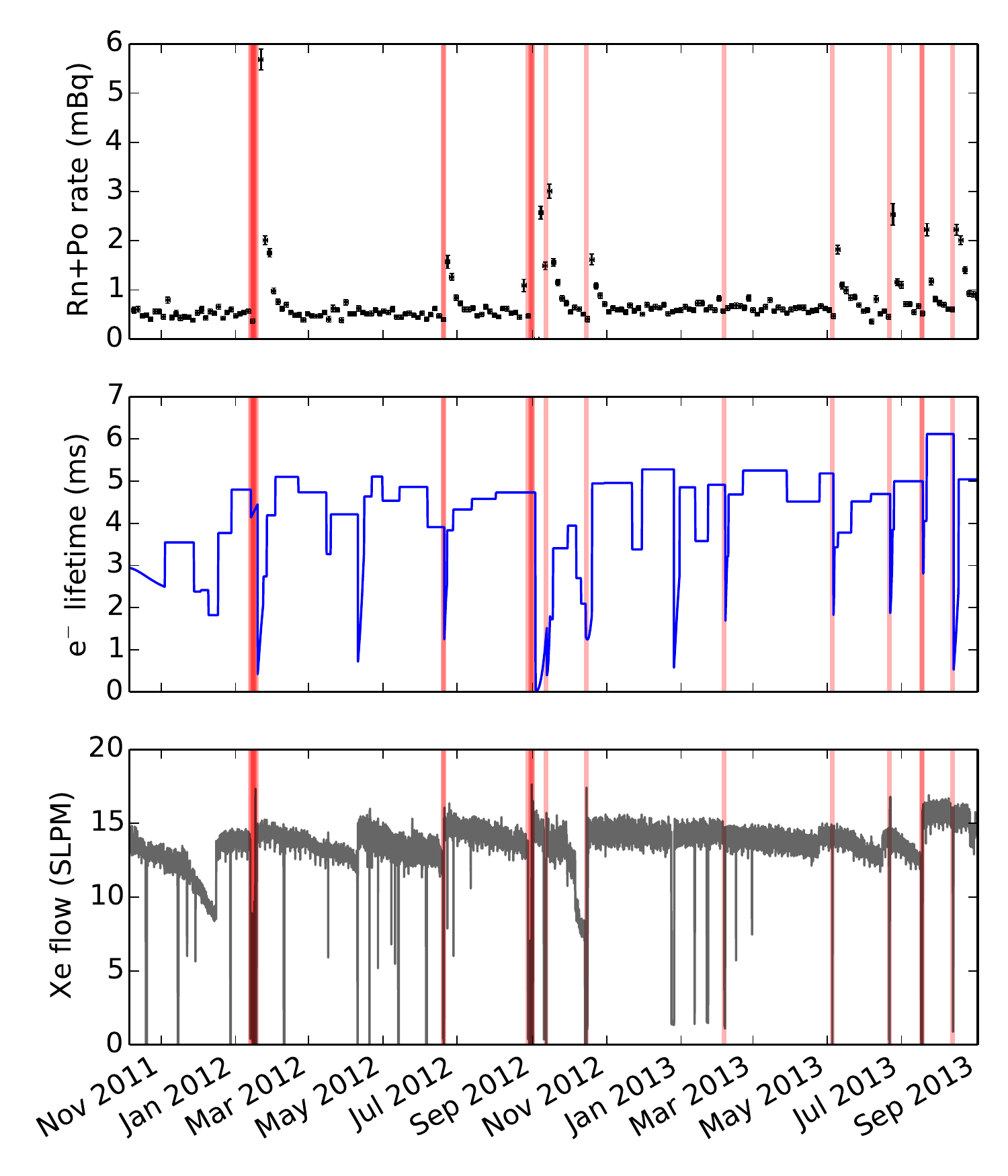}
\caption{(Color online) Top: rate of alpha decays in the region defined by Fig.~\ref{fig:recon_light_charge_alphas} (A), primarily from $^{222}\mathrm{Rn}$ and $^{218}\mathrm{Po}$ decays.
Middle: average electron lifetime measured in the TPCs.
Bottom: xenon recirculation flow rate.
Red lines indicate when xenon feed events occurred, wider lines correspond to larger feeds.
%Other dips in the purity are due to interruptions of the recirculation pump.
}
\label{fig:purity_plot}
\end{figure}

The rate of alpha decays of $^{222}\mathrm{Rn}$ and $^{218}\mathrm{Po}$ from the region A highlighted in Fig.~\ref{fig:recon_light_charge_alphas} (upper left), as a function of time spanning the data set, is shown in Fig.~\ref{fig:purity_plot}, along with the average electron lifetime and the xenon recirculation flow.
The rate of $^{222}\mathrm{Rn}$ and $^{218}\mathrm{Po}$ alpha decays points to two sources: the constant emanation from some component(s) in the recirculation loop, and injection from the high pressure xenon gas supply which is introduced during xenon feed events.

\subsection{Counting alpha decays} %$^{\bf{222}}\mathrm{\bf{Rn}}$ 
\label{sec:rates}
While fully reconstructed alpha events have good scintillation energy resolution, the V-wire detection efficiency is low ($<30\%$) and not well characterized at such small charge energies. 
Thus, for identifying alpha decays in the bulk LXe, only the scintillation and U-wire signals are used.
The detection efficiency is $>99.95\%$ for U-wire and APD channels for $^{222}\mathrm{Rn}$ decays (the lowest energy alpha decay considered). 

Alpha like events are selected if the event contains at least one scintillation signal between 25k and 66k counts.
Then the event then must pass one of three criteria: (a) the event has a single charge cluster and a single scintillation cluster (i.e.\ $^{222}\mathrm{Rn}$ or $^{218}\mathrm{Po}$), (b) the event has two scintillation clusters (Bi-Po), or (c) the event has one or more charge cluster(s) collected from the Bi decay before the only scintillation cluster (Po) (i.e. Bi scintillation is below threshold).
For Bi-Po events (b and c) a time window cut is applied which only selects events that are within $120$ and $800~\mathrm{\mu s}$ apart.
This ensures that all charge clusters from the first decay (Bi) have drifted out of the TPC before the second decay (Po) occurs, and that the charge cluster of the second decay is collected prior to the end of the waveform.
The efficiency of counting Bi-Po events in this time window is $55.7 \pm 7.2\%$, which is the fraction of exponential decay in the time window with a small correction due to events of type (c) which have an uncertainty due to lacking the Bi scintillation signal.
This factor was determined by applying the event selection criteria to Monte Carlo events with realistic thresholds and noise.

\begin{figure}
\includegraphics[width=0.48\textwidth]{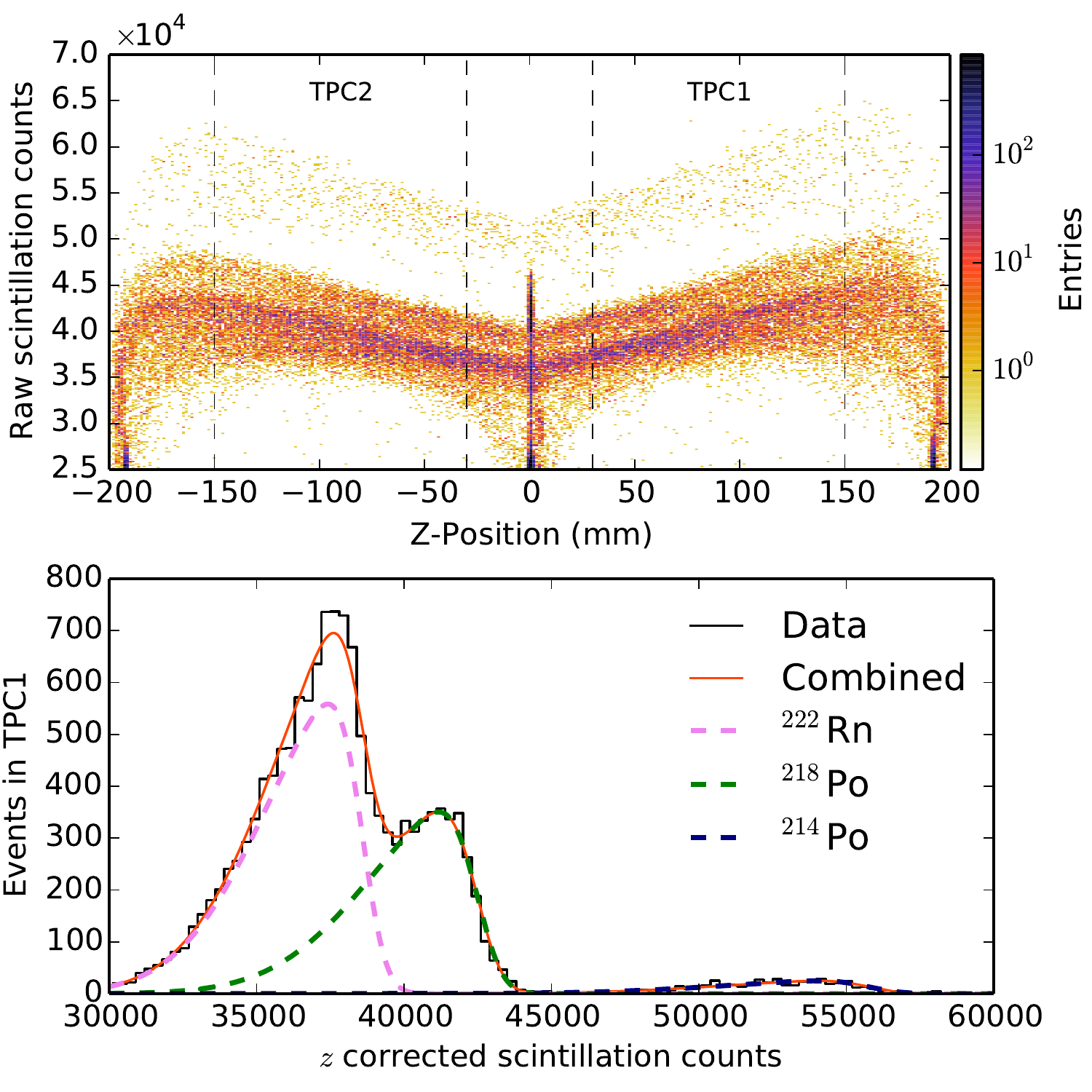}
\caption{(Color online) Top: raw scintillation signal $z$ position dependence for $^{222}\mathrm{Rn}$, $^{218}\mathrm{Po}$, and $^{214}\mathrm{Po}$ alpha decays. 
Vertical lines designate range of the data used ($30<|\mathrm{z}|<150~\mathrm{mm}$).
Bottom: histogram of the $z$-corrected scintillation signals in TPC1 with fits to each decay.}
\label{fig:scint_zpos_depend}
\end{figure}

In Fig.~\ref{fig:scint_zpos_depend} (top), the raw scintillation counts of alpha events found in this manner are plotted versus the $z$ position. 
It is seen that the alpha decays of $^{222}\mathrm{Rn}$, $^{218}\mathrm{Po}$, and $^{214}\mathrm{Po}$ are partially resolved.
The resolution of the raw scintillation is degraded towards the anodes and the cathode, where $x$-$y$ position dependent effects are more significant.
A linear $z$-dependent correction is applied to events in the range $30<|\mathrm{z}|<150~\mathrm{mm}$ in each TPC.
The resulting histogram of these events is shown in Fig.~\ref{fig:scint_zpos_depend} (bottom) for TPC1 (with a similar histogram for TPC2).
%After this correction, the $^{222}\mathrm{Rn}$ and $^{218}\mathrm{Po}$ events are partially resolved.
These peaks are then fit with three left-skewed Gaussian functions with a common skew factor and widths constrained by the ratio of their median energies.
%[0]*exp(-.5*((x-[1])/[2])^2)*Erfc([3]*(x-[1])/(sqrt(2)*[2])) +[4]*exp(-.5*((x-[5])/([5]/[1]*[2]))^2)*Erfc([3]*(x-[5])/(sqrt(2)*[5]/[1]*[2])) +[6]*exp(-.5*((x-[7])/([7]/[1]*[2]))^2)*Erfc([3]*(x-[7])/(sqrt(2)*[7]/[1]*[2]))}.
The resulting fits are integrated to determine the number of events associated with each decay in the analysis region (shown in Table~\ref{tab:counts}).
The parameter errors reported by the fit are used to establish the uncertainty on the number of decays of each alpha.
%Systematics are considered later in the analysis that uses these numbers.
\begin{table}
\caption{\label{tab:counts}Number of decays determined by the fits of the scintillation signals, per TPC, in the range $30<|z|<150~\mathrm{mm}$. Note the efficiency for counting $^{214}\mathrm{Po}$ is $55.7\pm7.2\%$.}
\begin{ruledtabular}
\begin{tabular}{lcc}
Decay & TPC1 & TPC2\\
\hline
$^{222}\mathrm{Rn}$ & $8036 \pm 189$ & $8069 \pm 188$\\
$^{218}\mathrm{Po}$ & $5549 \pm 106$ & $5286 \pm 110$\\
$^{214}\mathrm{Po}$ & $518 \pm 23$ & $521 \pm 23$\\
\end{tabular}
\end{ruledtabular}
\end{table}
The left skewed Gaussian function was chosen based on the best overall $\chi^2$, after also considering Gaussian and Crystal Ball functions.
%This skew in the raw scintillation is believed to be due to radial dependent effects of the light collection efficiency. 
%[0]*exp(-.5*((x-[1])/[2])^2)*Erfc([3]*(x-[1])/(sqrt(2)*[2])) +[4]*exp(-.5*((x-[5])/([5]/[1]*[2]))^2)*Erfc([3]*(x-[5])/(sqrt(2)*[5]/[1]*[2])) +[6]*exp(-.5*((x-[7])/([7]/[1]*[2]))^2)*Erfc([3]*(x-[7])/(sqrt(2)*[7]/[1]*[2]))

The number of bulk events associated with subsequent decays in the series is seen to decrease. 
This is because each decay in the chain has a probability of producing a positively ionized daughter that drifts towards the cathode and out of the analysis volume.
The fraction of ions produced by each decay is discussed in Section~\ref{sec:alpha_ion_fraction} and Section~\ref{sec:beta_ion_fraction}.

\subsection{$^{\bf{222}}\mathrm{\bf{Rn}}$ - $^{\bf{218}}\mathrm{\bf{Po}}$ coincidences}
Using both the spatial and temporal information provided by the TPC, a $^{218}\mathrm{Po}$ decay can often be associated with its parent $^{222}\mathrm{Rn}$ decay.
Such delayed coincidences provide information about the $^{218}\mathrm{Po}$ charge state and the motion of individual daughter atoms or ions in LXe.

\begin{figure}
\includegraphics[width=0.48\textwidth]{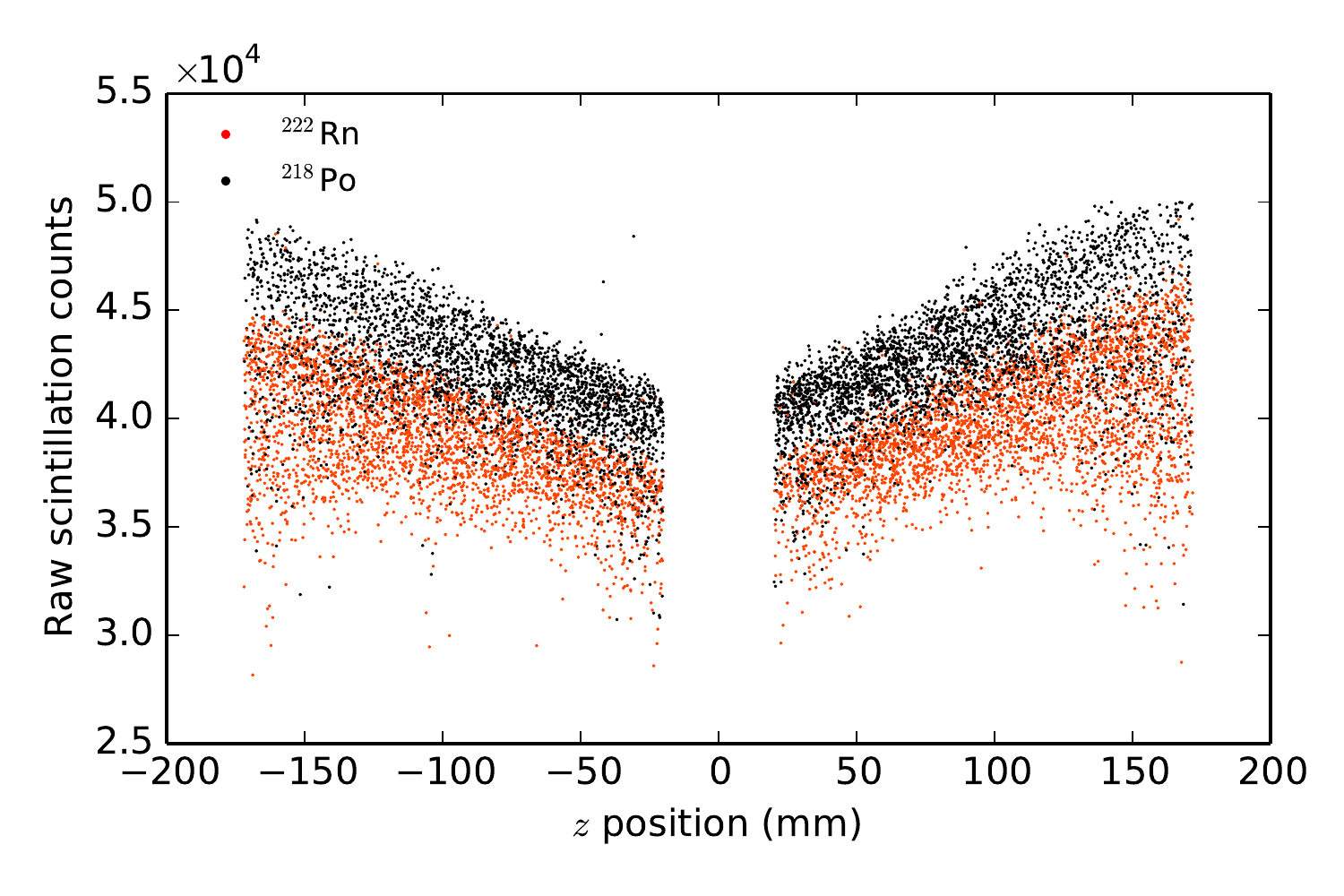}
\caption{(Color online) Raw scintillation counts versus $z$ position of coincident pairs assigned as $^{222}\mathrm{Rn}$ (red) and $^{218}\mathrm{Po}$ (black) in the region $20 < |z| < 172~\mathrm{mm}$.
}
\label{fig:coincidence_scint_zdep}
\end{figure}

Events used to search for $^{222}\mathrm{Rn}$-$^{218}\mathrm{Po}$ (Rn-Po) coincidences have a single charge cluster, a raw scintillation signal between 28k and 50k counts, and are between $20 <|z|< 172~\mathrm{mm}$.
The pairing algorithm requires that events occur within $3$ minutes of each other and have charge collected on the same or on an adjacent U-channel, allowing for small transverse motion due to diffusion or xenon motion.
In the case of two fully reconstructed events, they are also required to be detected by the same or an adjacent V-wire channel. 
Alpha candidate events that have more than one possible coincidence match are removed from the analysis ($< 1\%$ of the total).
Coincidence events are assigned as $^{222}\mathrm{Rn}$ and $^{218}\mathrm{Po}$ decays based on their order in time. 
This technique is possible because of the very low rate of alpha decays in the detector. 
Fig.~\ref{fig:coincidence_scint_zdep} shows the raw scintillation signal versus the $z$ position of all paired Rn and Po events. 
The energy differences between Rn and Po populations corroborate the decay assignment based on time ordering.
The total number of Rn-Po pairs found in the data set is $6507$, with $277$ of these having both decays fully reconstructed and $2083$ with one decay fully reconstructed. 

The EXO-200 Geant4-based Monte Carlo \cite{Albert2014_2vbb} was used to determine the probability that two alpha decays randomly placed in the LXe would have their charge collected on the same or adjacent U-wire and therefore be found in coincidence using the pairing method described. 
This simulation has a detailed model of the detector, including the wire channel layout, and determined that the spatial requirement rejects $95.9\%$ of random (false) coincidences.

The rate of false coincidences can also be estimated from the data, using the fully reconstructed candidate events. 
For this data set, matching done with only the U-channel information finds an additional 16 coincidences.
Further investigation of those $16$ pairs identifies $3$ with separations in the $x$-$y$ plane of $>100~\mathrm{mm}$, while the remaining $13$ events have light map corrected scintillation energies and separations in the $x$-$y$ plane ($<30~\mathrm{mm}$) that are compatible with valid coincidences. 
The number of false coincidences is therefore estimated at $<2\%$ for the Rn-Po pairs matched using U-wires only. 

\begin{figure}
\includegraphics[width=0.48\textwidth]{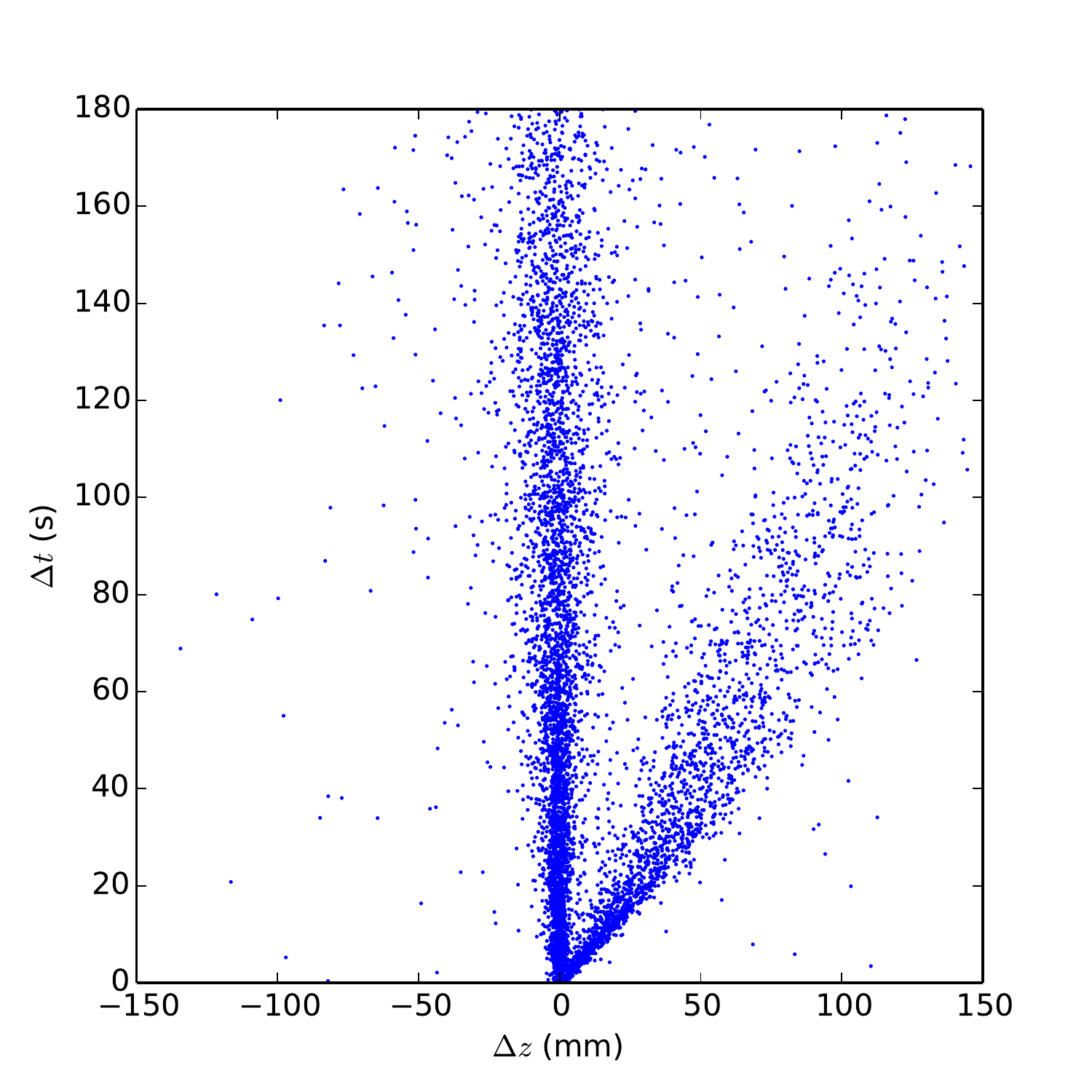}
\caption{(Color online) Scatter plot of $^{218}\mathrm{Po}$ drift distance versus time between the $^{222}\mathrm{Rn}$ and $^{218}\mathrm{Po}$ decays. Displacement ($\Delta z$) is defined as positive when movement is towards the cathode.}
\label{fig:coincidence_2D_z_t}
\end{figure}

\begin{figure}
\includegraphics[width=0.48\textwidth]{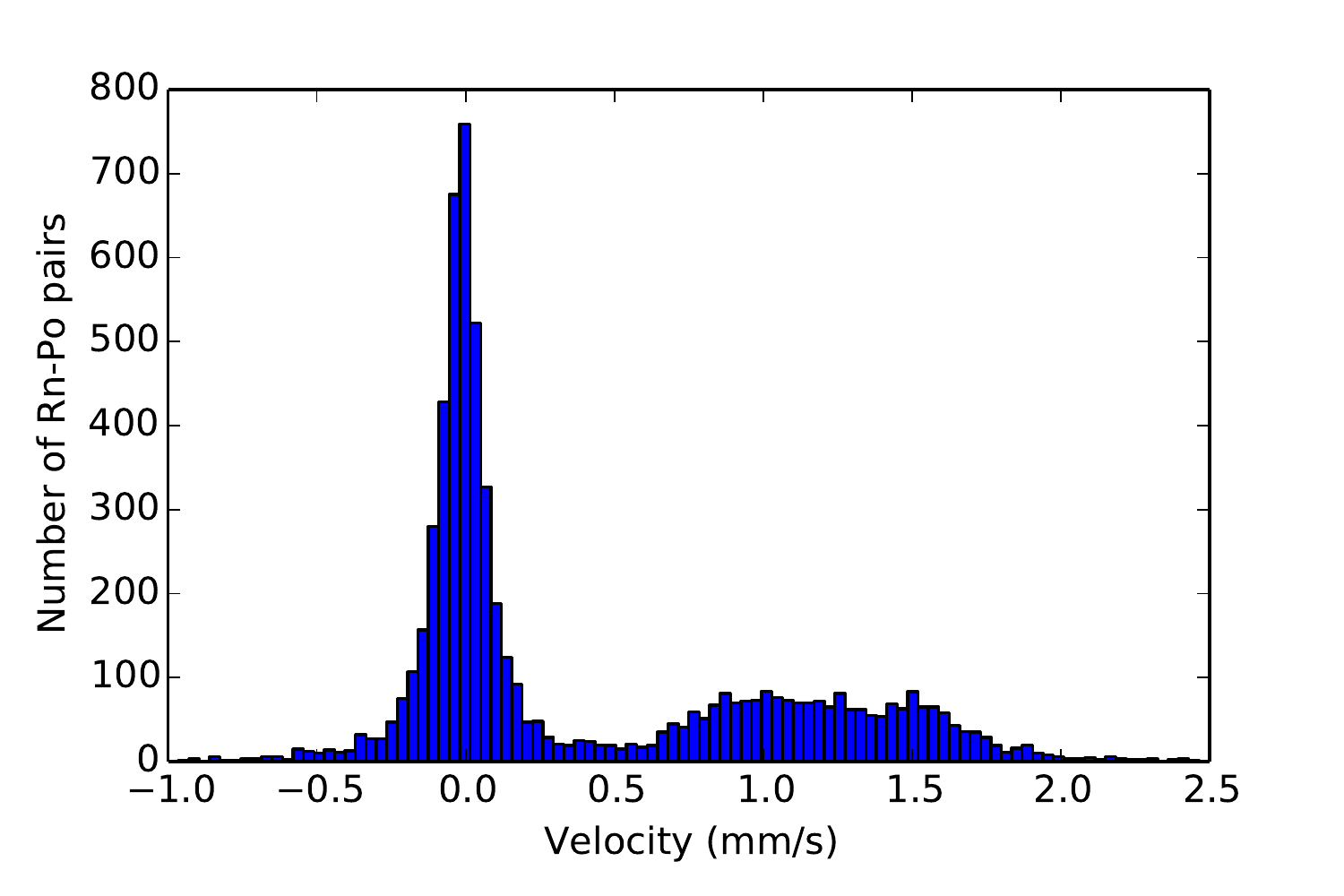}
\caption{(Color online) Histogram of the mean velocity of $^{218}\mathrm{Po}$ ions and atoms extracted from $^{222}\mathrm{Rn}$-$^{218}\mathrm{Po}$ coincidences.}
\label{fig:coincidence_velocity}
\end{figure}

The position difference ($\Delta z$) versus the time difference ($\Delta t$) for these Rn-Po coincidences is shown in Fig.~\ref{fig:coincidence_2D_z_t}, where $\Delta z$ is defined as positive for a displacement towards the cathode. 
Two primary populations are observed which are attributed to $^{222}\mathrm{Rn}$ that: (a) produced neutral $^{218}\mathrm{Po}$ atoms which only moved slightly, and (b) produced positive $^{218}\mathrm{Po}$ ions which drifted towards the cathode. 
A histogram of the mean drift velocity of these $^{218}\mathrm{Po}$ ions and atoms is shown in Fig.~\ref{fig:coincidence_velocity}.
The mean velocity of $^{218}\mathrm{Po}$ ions is significantly broader than for the $^{218}\mathrm{Po}$ neutrals. 
The cause for this broadening of ions is discussed in section~\ref{sec:ion_model}.
 
Matching $^{218}\mathrm{Po}$ and $^{214}\mathrm{Po}$ decays has also been attempted with the method described here.
Due to the longer decay time, accurate assignment is difficult without improvements to the V-wire detection threshold and/or to the matching algorithm.

%%%%%%%%%%%%%%%%%%%%%%%%%%%%%%%%%%%%%%%%%%%%%%%%%%%%%%%%%%%%%%%%%%%%%%%%%%%%%%%%%%%%%%%
\section{Results and discussion}
%%%%%%%%%%%%%%%%%%%%%%%%%%%%%%%%%%%%%%%%%%%%%%%%%%%%%%%%%%%%%%%%%%%%%%%%%%%%%%%%%%%%%%%

\subsection{Liquid xenon flow}
\begin{figure}
\includegraphics[width=0.48\textwidth]{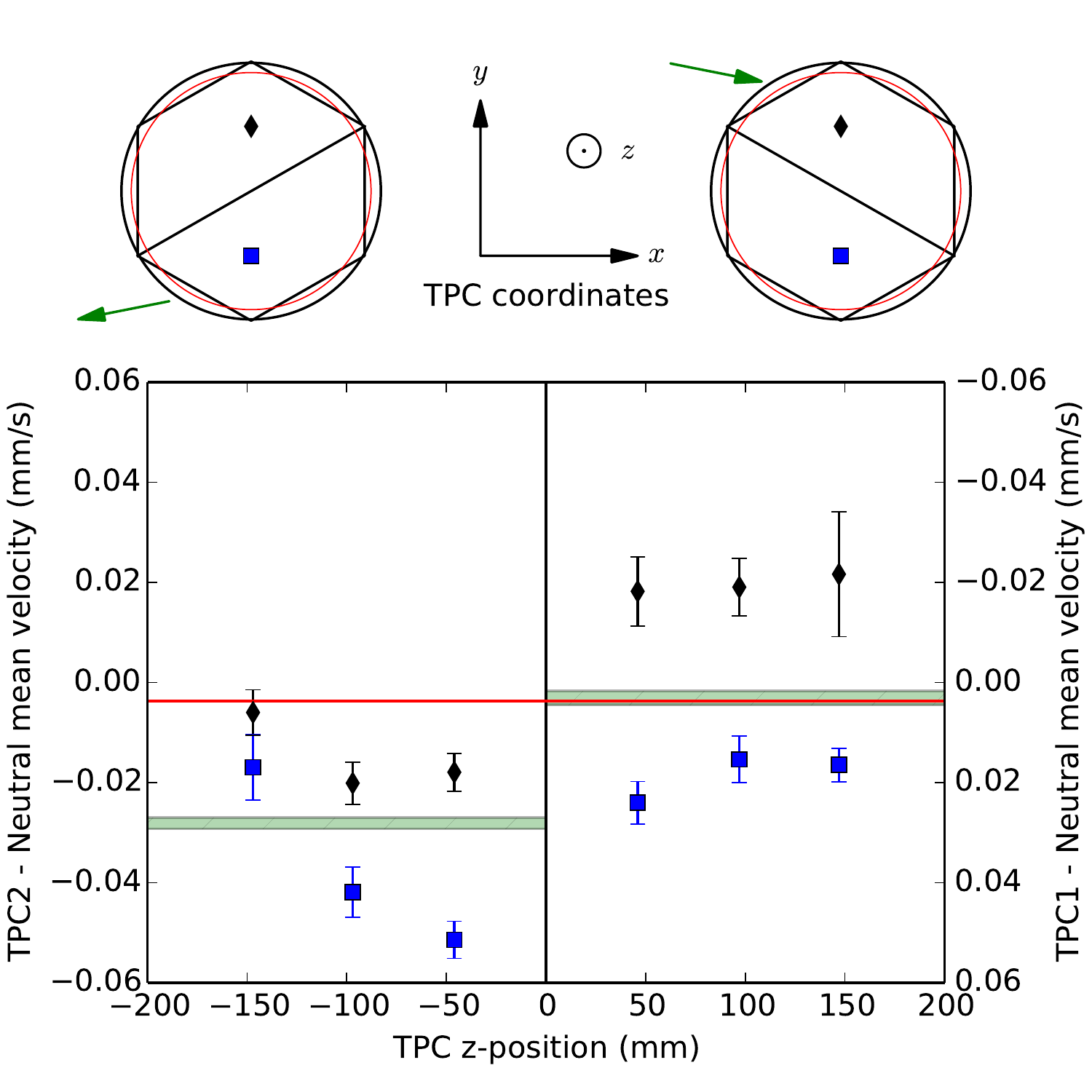}
\caption{(Color online) Top sketch: the ``upper'' and ``lower'' volumes in each TPC are defined using the U-wires as shown in the hexagonal active volume. 
Black circle shows the internal diameter of the detector, and red circle shows the position of the teflon reflector. 
Green arrows show where the LXe flows into and out of the detector at the end caps, near the anodes.
Bottom: the mean velocity of neutral $^{218}\mathrm{Po}$, measured in the different volumes of the EXO-200 detector.
The green band shows the average velocity of all events in each TPC, and the red line shows the expected velocity due to the LXe recirculation.
Note that the vertical axis on TPC1 is reversed to account for the convention on the sign of the velocity in which velocities are defined as positive for motion towards the cathode.}
\label{fig:neutrals_xenon_flow}
\end{figure}

The flow of LXe in the detector can be measured using the Rn-Po coincidence events that produce neutral $^{218}\mathrm{Po}$ atoms.
The detector is divided into 12 equal volumes, with 6 bins in $z$ and 2 bins in the transverse direction using the U-wire position.
Velocity histograms of the neutral pairs in each volume are fit with a Gaussian to determine the mean velocity and error.
The results, shown in Fig.~\ref{fig:neutrals_xenon_flow}, indicate non-uniform flow within the detector.
%LXe is also able to flow around the TPCs which can create flow patterns and allows these large velocities. 
The observed mean velocity of $^{218}\mathrm{Po}$ atoms in the active volume of TPC1 is $1.9\pm 2.6\times 10^{-3}~\mathrm{mm/s}$, while the observed mean velocity in the active volume of TPC2 is $-26.4\pm 2.3\times 10^{-3}~\mathrm{mm/s}$ (green bands). 
Note that, because of our convention for $\Delta z$ and $v$ in Figs.~\ref{fig:coincidence_2D_z_t} and \ref{fig:coincidence_velocity}, these numbers imply motion in the same direction.
For comparison, a velocity of $3.6 \times 10^{-3}~\mathrm{mm/s}$ in the $-z$ direction would be obtained for a recirculation rate of $14~\mathrm{SLPM}$ (gas), assuming that the flow was uniform across the $\approx 20~\mathrm{cm}$ detector radius (red line). 
The average motion is aligned with the direction of LXe recirculation, but it is quite non-uniform, implying substantial flow outside of the active regions of the TPCs (i.e. outside the teflon reflector or hexagonal wire plane regions).  
Note that the average flow direction is the same in the bottom of both TPCs, but it is in opposing directions in the top regions of the two TPCs.
In either case fluid flow is significantly less than the ion velocity (of order $1~\mathrm{mm/s}$).

\subsection{$^{\bf{218}}\mathrm{\bf{Po}}$ ion drift velocity and mobility}
\label{sec:ion_model}
\begin{figure}
\includegraphics[width=0.48\textwidth]{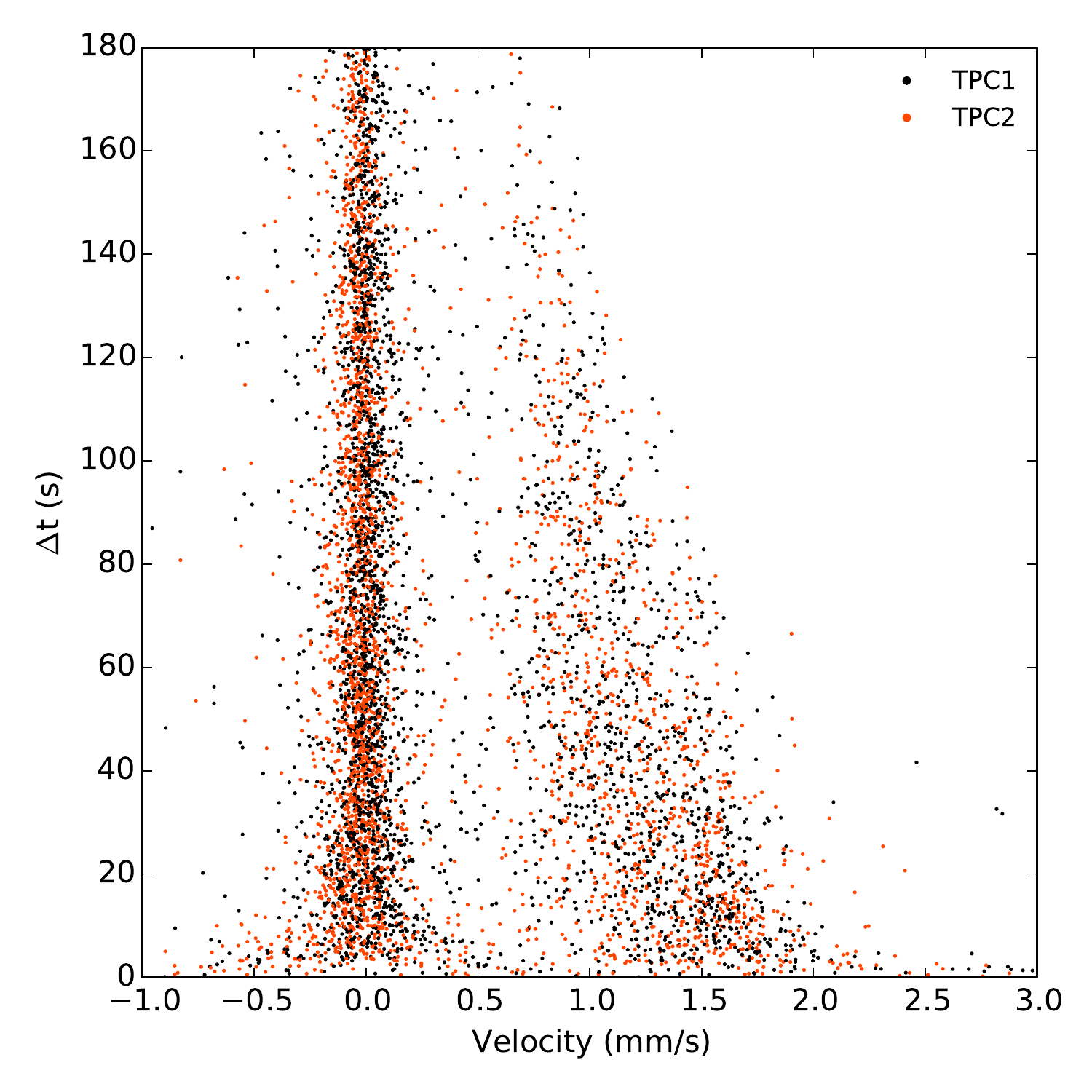}
\caption{(Color online) The velocity versus time between Rn-Po coincident events (positive is towards the cathode). 
The TPC in which the events occur is indicated by the color of the points.}
\label{fig:coincidence_2D_v_t}
\end{figure}

The broadening of the velocity distribution of positive $^{218}\mathrm{Po}$ ions relative to that of neutral $^{218}\mathrm{Po}$ atoms in Fig.~\ref{fig:coincidence_velocity} can be understood  by plotting the drift time ($\Delta t$) versus velocity, as shown in Fig.~\ref{fig:coincidence_2D_v_t}.
It is seen that the average ion velocity decreases with $\Delta t$.
A possible explanation for this effect is that the $^{218}\mathrm{Po}$ ion initially moves with a higher drift velocity $v_1$, and while drifting a reaction or charge transfer occurs resulting in a larger molecular ion or a reduction of the charge of the ion.
After this reaction the ion moves with a lower drift velocity, $v_2$.
Since the velocity is measured using the initial and final decay positions and times, the observed velocity is the time average spent at $v_1$ and $v_2$. 

\begin{figure}
\includegraphics[width=0.48\textwidth]{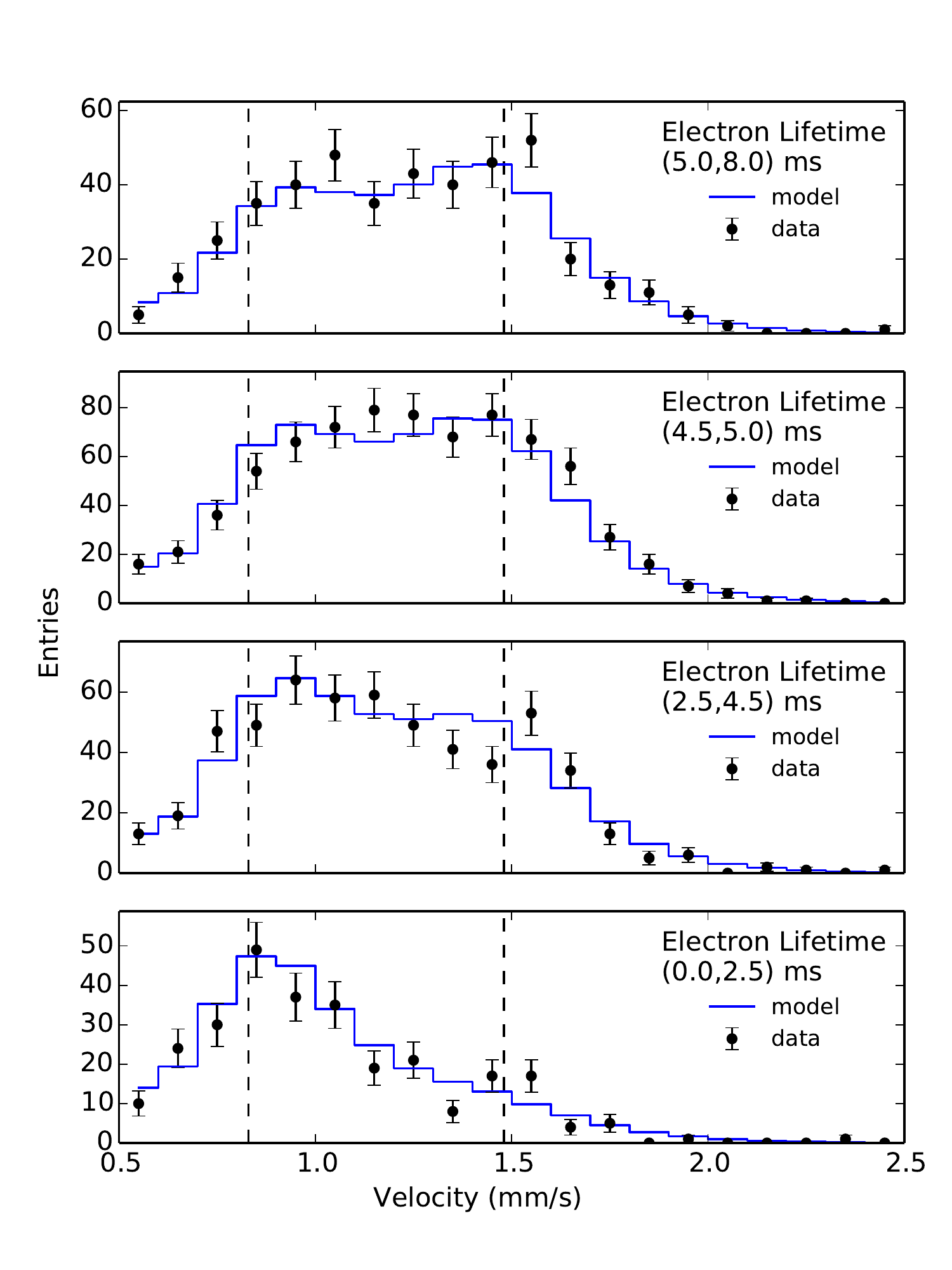}
\caption{(Color online) Histograms of the average velocity of positive ions separated into four ranges of electron lifetimes. 
A model (solid lines) is used to fit the data (points with error bars).
The dashed lines indicate the velocities $v_1$ and $v_2$ of the fit.}
\label{fig:velocity_purity_bins}
\end{figure}

It is interesting to compare the velocity distribution for data sets with different ranges of electron lifetime ($\tau_e$).
In Fig.~\ref{fig:velocity_purity_bins} it is observed that the ion velocity distributions shift towards lower values as $\tau_e$ decreases, suggesting that the presence of impurities affects the ion transport. 
A Monte Carlo model with five input parameters ($v_1$, $v_2$, $C$, $N$, $D$) is used to describe the ion velocity data shown in Fig.~\ref{fig:velocity_purity_bins}.
In addition to $v_1$ and $v_2$ (described previously), the parameters $C$ and $N$ are the ratios of the ion reaction and neutralization time constants to $\tau_e$, and $D$ is an effective 1-dimensional diffusion constant.
To match the actual $\tau_e$ distribution in the data, the $\tau_e$ in each simulation is randomly selected from the data set of $\tau_e$ values of all Rn-Po coincidences.
The initial charge state of the $^{218}\mathrm{Po}$ is randomly chosen with probabilities that match the observed ratio of ions to atoms (see Section~\ref{sec:alpha_ion_fraction}).
%The $\tau_e$ in each simulation is randomly selected from this set.
The simulation then generates a random decay time ($\Delta t$), reaction time, and neutralization time from their respective exponential distributions, and the initial $z$ position from a uniform distribution.
The $\Delta z$ of the ion drift is then calculated.
Diffusion and an additional term, $\delta z$, representing the $1~\mathrm{mm}$ uncertainty of $z$ position (for details see \cite{Albert2014_2vbb}) is added to the final position as a normally distributed offset with variance $2 D \cdot \Delta t + \delta z^2$.
The Monte Carlo generates 20-million simulations, which are histogrammed and scaled by the ratio of coincidences to simulations.
The model parameters are found by minimizing the combined $\chi^2$ of all four histograms, and uncertainties are found by manually profiling each variable at this minimum.
The minimum has a $\chi^2$ of $73.9$ with $63$ degrees of freedom.
The best fit parameters values with 1-sigma errors are
$v_1=1.48 \pm 0.01 ~\mathrm{mm/s}$,
$v_2=0.83 \pm 0.01 ~\mathrm{mm/s}$,
$C=12600 \pm 660$,
$N = 6.0_{-1.7}^{+4.9}\times 10^{5}$,
$D=0.61\pm 0.04 ~\mathrm{mm^2/s}$.

The value of $D$ is two orders of magnitude larger than the expected diffusion coefficient of an ion in LXe, so this parameter presumably represents the effect of xenon motion rather than proper diffusion.
The agreement between this model and data supports the reaction model with a reaction time constant proportional to $\tau_e$.
Another scenario to consider is that the initial species is $^{218}\mathrm{Po}^{++}$ and the reaction is to $^{218}\mathrm{Po}^{+}$.
This charge transfer could be aided by an impurity of lesser ionization potential than $^{218}\mathrm{Po}^{+}$.
%However, the initial existence of only charge zero and charge $2+$ daughters in this scenario seems unlikely.
However, this would require the survival of only charge $0$ and $2+$ daughters after the decay, with no significant fraction of $1+$ daughters, an a-priori unlikely scenario.
The effect of charge-induced fluid motion by the drifting xenon holes has also been considered, but estimates indicate that it is negligible.  
Using the drift electric field in the detector, $380 \pm 5~\mathrm{V/cm}$, the mobilities of the two drifting species are
$0.390 \pm 0.006~\mathrm{cm}^2/(\mathrm{kV}~\mathrm{s})$ 
and 
$0.219 \pm 0.004~\mathrm{cm}^2/(\mathrm{kV}~\mathrm{s})$,
for $v_1$ and $v_2$, respectively.
The slower mobility ($v_2$) is similar to mobility measurements of other atomic ions, while the initial mobility ($v_1$) is significantly higher than any ion mobility measurements found in the literature (with the exception of xenon holes, which are understood to be $\mathrm{Xe}^{2+}$ ions with charge motion predominantly by resonant charge transfer \cite{hilt1994positive}).

\subsection{Alpha decay: $^{\bf{218}}\mathrm{\bf{Po}}$ ion fraction }
\label{sec:alpha_ion_fraction}
\begin{figure}
\includegraphics[width=0.48\textwidth]{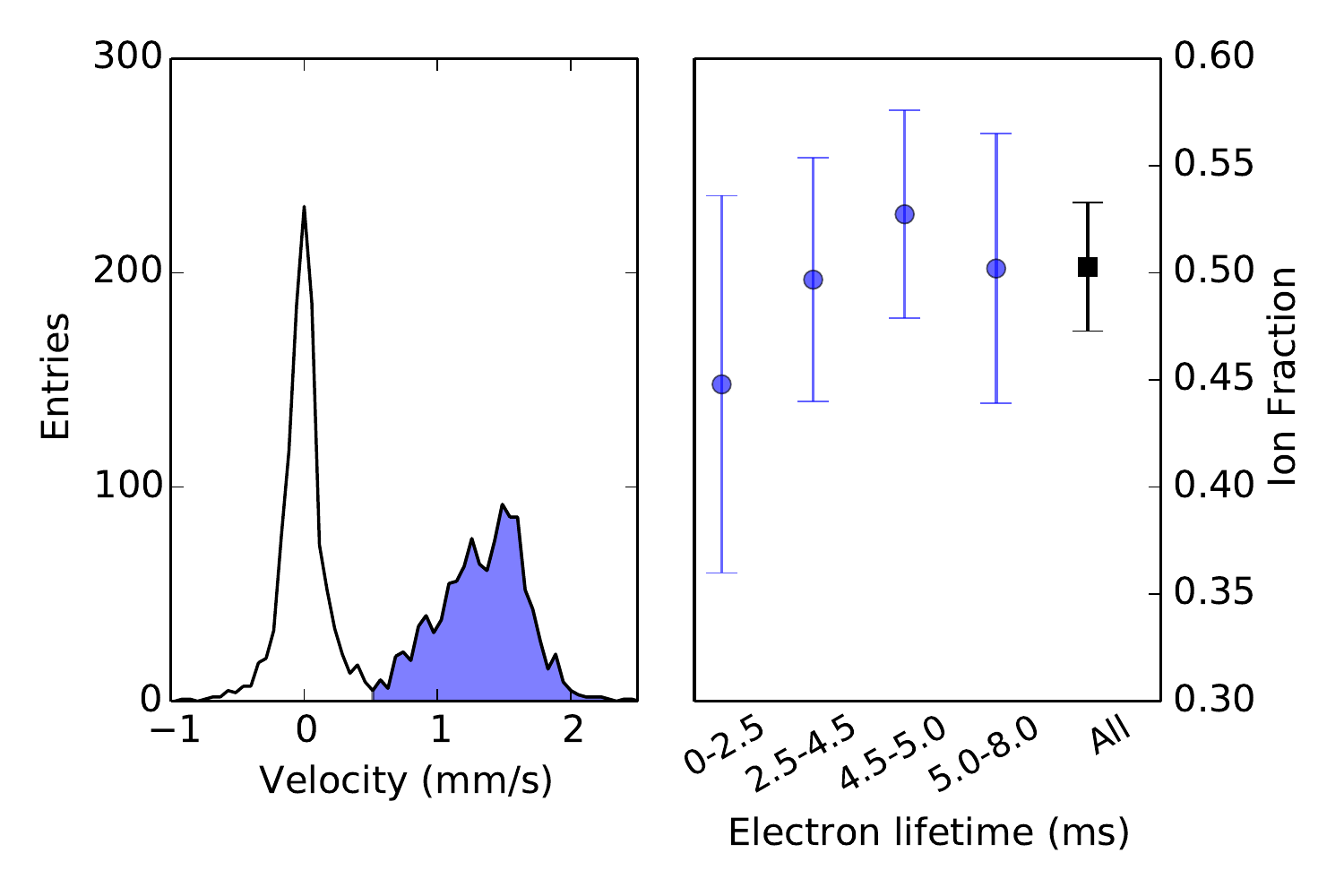}
\caption{(Color online) Left: velocity histogram after applying the $\Delta t$ cut on all of the Rn-Po coincidences. 
The shaded region (velocity $>$ 0.5 mm/s) is integrated to determine the number of $^{218}\mathrm{Po}$ ions.
Right: the ion fraction found for different electron lifetime ranges.}
\label{fig:coincidence_ion_fraction}
\end{figure}

Using the Rn-Po coincidences it is possible to calculate the fraction of $^{222}\mathrm{Rn}$ alpha decays that produce ionized $^{218}\mathrm{Po}$ ($f_{\alpha}$).
To have an equal probability of selecting neutral and ion daughters, a cut on $\Delta t$ is applied to ensure that no ions can drift outside the analysis volume.
This cut is $\Delta t < |z_\mathrm{Rn}|/v_m$, where $z_\mathrm{Rn}$ is the position of the Rn decay and $v_m$ is a conservative maximum velocity of $2.5~\mathrm{mm/s}$.
Fig.~\ref{fig:coincidence_ion_fraction} (left) shows the velocity distribution of all coincidences passing this unbiased cut. 
The ion fraction is the number of ions (entries with velocity between $0.5~\mathrm{mm/s}$ and $2.5~\mathrm{mm/s}$) divided by the total number of entries. 
As shown in Fig.~\ref{fig:coincidence_ion_fraction} (right), $f_{\alpha}$ is independent of the electron lifetime within statistical uncertainties.
%Error on the estimated number of ions is calculated using the standard binomial formulation:
%\[ 
%\sigma_k = \sqrt{N \epsilon \left( 1- \epsilon \right)} .
%\] 
It is found that $f_\alpha = 50.3 \pm 3.0\%$, where the error is statistical, and the uncertainty on the separation of the neutral and ion populations is negligible.
To our knowledge this is the first measurement of the daughter ion fraction of an alpha decay in a noble liquid.

\subsection{Beta decay: $^{\bf{214}}\mathrm{\bf{Bi}}$ ion fraction} 
\label{sec:beta_ion_fraction}
The ion fraction of  $^{214}\mathrm{Bi}$ ($f_{\beta}$), the daughter of $^{214}\mathrm{Pb}$ beta decay, can be estimated using the observed activity of the $^{218}\mathrm{Po}$ and $^{214}\mathrm{Po}$ decays in the analysis volume (see Table~\ref{tab:counts}).
As seen in Fig~\ref{fig:Rn222_decay_chain}, the chain following $^{218}\mathrm{Po}$ is 
%{^{222}\mathrm{Rn}} \xrightarrow{\alpha} {^{218}\mathrm{Po}} \xrightarrow{\alpha}
\[
{^{218}\mathrm{Po}} \xrightarrow{\alpha} {^{214}\mathrm{Pb}} 
\xrightarrow{\beta} \left( {^{214}\mathrm{Bi}}\cdot{^{214}\mathrm{Po}} \right)
\xrightarrow{\alpha} {^{210}\mathrm{Pb}}.
\]
Here the $^{214}\mathrm{Bi}$ and $^{214}\mathrm{Po}$ decays are lumped together because negligible ion drift occurs during the short $^{214}\mathrm{Po}$ half life. 

Since the $^{214}\mathrm{Pb}$ ions are swept to the cathode in a time that is short compared with their half life, the main component for the decays in bulk LXe originates from the $^{218}\mathrm{Po}$ decays that result in neutral atoms.
This is expressed as $1-f_\alpha'$.
In addition, the fraction of ionic $^{214}\mathrm{Pb}$ that decay while drifting, $\epsilon_{\mathrm{Pb}}$, gives a contribution $f_\alpha' \cdot \epsilon_\mathrm{Pb}$ to the measured number of $^{214}\mathrm{Pb}$ decays.
The ion fraction, $f_\alpha'$, from the alpha decay of $^{218}\mathrm{Po}$, may be different from $f_\alpha$, the ion fraction from alpha decays of $^{222}\mathrm{Rn}$.
However, it is reasonable to approximate $f_\alpha'$ with the value of $f_\alpha$ measured in Section~\ref{sec:alpha_ion_fraction} since the decay energies are similar and the ionization potentials ($8.41~\mathrm{eV}$ for Po  and $7.41~\mathrm{eV}$ for Pb) only differ by $1~\mathrm{eV}$.
The factor $\epsilon_{\mathrm{Pb}} = 0.088 \pm 0.023$ is calculated using the Monte Carlo model and parameters in section~\ref{sec:ion_model}, with the half life of $^{218}\mathrm{Po}$ and an initial $z$ distribution of $^{218}\mathrm{Po}$ calculated from a previous simulation of Rn-Po drift, which started with a uniform $z$ distribution of $^{222}\mathrm{Rn}$ from the anode to the cathode.
The number of $^{214}\mathrm{Po}$ decays occurring in the bulk LXe is similarly related to the number of $^{214}\mathrm{Pb}$ events, with $f_{\beta}$ as the ion fraction of $^{214}\mathrm{Bi}$ and ion contribution factor $\epsilon_{\mathrm{Bi}} = 0.108 \pm 0.028$ calculated with another iteration of simulations of $^{214}\mathrm{Pb}$.
The ratio of $^{214}\mathrm{Po}$ events ($A^{214}_{\mathrm{Po}}$) to $^{218}\mathrm{Po}$ events ($A^{218}_{\mathrm{Po}}$) in the bulk, is therefore equal to 
\[
\frac{A_{\mathrm{Po}}^{214}}{A_{\mathrm{Po}}^{218}}= (1-f_{\alpha}' + f_{\alpha}'\epsilon_{\mathrm{Pb}}) (1-f_{\beta} + f_{\beta}\epsilon_{\mathrm{Bi}}).
\]
From Table~\ref{tab:counts} and the $^{214}\mathrm{Po}$ efficiency factor, $A_\mathrm{Po}^{214}/A_{\mathrm{Po}}^{218} = 0.172 \pm 0.023$.
In addition to the statistical uncertainties in Table~\ref{tab:counts}, the difference of the measured ratios in the two TPCs and the mean was included as a gauge of systematic errors in the skewed Gaussian fits.
Under the assumption that $f_{\alpha}' = f_{\alpha}$, the equation above can be solved for $f_\beta$, finding $f_\beta = 76.4 \pm 5.7\%$.

The same prescription can be followed to relate the activity of $^{214}\mathrm{Po}$ to that of $^{222}\mathrm{Rn}$, with the addition of one more step ($^{222}\mathrm{Rn}$-$^{218}\mathrm{Po}$). 
The motivation for this is to check that the left-skewed Gaussian fit does not systematically increase either $^{218}\mathrm{Po}$ or $^{222}\mathrm{Rn}$ counts.
In this step, the $f_{\alpha}$ has been measured, and the ion contribution factor $\epsilon_{Po} = 0.344\pm 0.003$ is found using the Monte Carlo with a uniform initial $z$ distribution of $^{222}\mathrm{Rn}$.
The expression relating these two activities is
\[
\frac{A_{\mathrm{Po}}^{214}}{A_{\mathrm{Rn}}^{222}}= (1-f_{\alpha} + f_{\alpha}\epsilon_{\mathrm{Po}}) (1-f_{\alpha}' + f_{\alpha}'\epsilon_{\mathrm{Pb}}) (1-f_{\beta} + f_{\beta}\epsilon_{\mathrm{Bi}}),
\]
where $A_{\mathrm{Po}}^{214} / A_{\mathrm{Rn}}^{222} = 0.116 \pm 0.016$.
With this procedure $f_{\beta}$ is determined to be $76.3 \pm 6.2\%$, which is in good agreement with the measurement starting from $^{218}\mathrm{Po}$.

Without considering the assumption that $f_\alpha\ = f_\alpha'$, the primary source of uncertainty on $f_{\beta}$ ($^{218}\mathrm{Po} \rightarrow ^{214}\mathrm{Po}$) is the $^{214}\mathrm{Po}$ counting efficiency, contributing $2/3$ of the total uncertainty.
Uncertainties on $\epsilon_{\mathrm{Pb}}$ and $\epsilon_{\mathrm{Bi}}$ are estimated by running the model with two extreme assumptions for the unknown velocity of these ions, using $v_1$ and $v_2$ found in Section~\ref{sec:ion_model} as limits.
The uncertainty on $\epsilon_{\mathrm{Po}}$ is calculated using the model by incorporating the uncertainties on the parameters $C$, $v_1$, and $v_2$.
Finally, treating the assumption that $f_{\alpha} \neq f_{\alpha}'$, in the limit that  $f_{\alpha}' = 0$ the maximum value of $f_{\beta}$ is $93\%$.
Alternatively, for $f_{\beta}$ to be $< 50\%$ would require that $f_{\alpha}'>76\%$.

\subsection{Alpha and beta ion and xenon hole production}
It is interesting to compare the alpha and beta decay daughter ion fractions to the surviving ion fraction of xenon holes, $f_h=1-r$, where $r$ is the fraction of initial ionizations that lead to recombination.  
For beta decay, recombination results in the generation of a photon, so the number of scintillation photons per ionization is $r+\alpha$, where $\alpha$ is the initial ratio of the number of excited xenon atoms, or excitons, to ionizations.  
The ratio of photons ($N_{ph}$) to electrons ($N_e$) from the decay is
\[
\frac{N_{ph}}{N_e} = \frac{\alpha+r}{1-r} 
\]
which gives
\[
f_{h}=1-r=\frac{1+\alpha}{1+\frac{N_{ph}}{N_e}}
\]
Using a mean experimental value for LXe, $\alpha=0.13$, \cite{Doke2002, Aprile2007} and $N_{ph}=30.5~\mathrm{keV}^{-1}$ and $N_e=42~\mathrm{keV}^{-1}$ at the EXO-200 drift field from NEST \cite{NEST} results of $^{214}\mathrm{Pb}$ decay, $f_h=0.65$ for holes from beta decay.  
In comparison, for $5.3~\mathrm{MeV}$ alphas, measurements give $f_h \approx 0.02$ \cite{Ichinose1991}. 
Since the $^{218}\mathrm{Po}$ daughter of $^{222}\mathrm{Rn}$ alpha decay should be located at the end of a $101~\mathrm{keV}$ nuclear recoil track, it is perhaps more appropriate to compare to the hole ion fraction for a recoil track of such energy, $f_h = 0.18$ \cite{mu2015ionization}.

The observed ion fraction for daughter ions is greater than that for Xe holes in both beta decay and alpha decay, as well as the nuclear recoil.
This suggests the existence of an additional reaction of the daughter atoms that competes with recombination, particularly in high density ionization tracks in LXe. 
Both charge transfer from holes to the $^{218}\mathrm{Po}$ daughter atom and Penning ionization of the daughter atom by excitons (excited Xe atoms) are energetically allowed in liquid xenon and could substantially increase the daughter ion fraction relative to that of holes in alpha decay or nuclear recoil.
These two processes could also contribute to enhancement of the daughter ion fraction in beta decay, although to a lesser extent due to the lower density of ionization and excitation.

\section{Conclusion}
Alpha decays provide unique signatures in the EXO-200 detector that are useful to directly studying a number of low-energy physical processes.
Detailed measurements of $^{222}\mathrm{Rn}$ and its decay products in the EXO-200 detector have been presented.

%A matching algorithm was described that associates $^{222}\mathrm{Rn}$ decays to their daughter $^{218}\mathrm{Po}$ decays.
Delayed coincidences between $^{222}\mathrm{Rn}$ and daughter $^{218}\mathrm{Po}$ alpha decays have been used to determine the fraction of $^{218}\mathrm{Po}$ ions that result from $^{222}\mathrm{Rn}$ alpha decay in LXe, $f_\alpha=50.3 \pm 3.0\%$, and to study the mobility of the daughter ion.
Two distinct ion mobilities were found: 
$0.390 \pm 0.006~\mathrm{cm}^2/(\mathrm{kV}~\mathrm{s})$, 
and 
$0.219 \pm 0.004~\mathrm{cm}^2/(\mathrm{kV}~\mathrm{s})$, 
with the ions initially moving at the higher mobility.
The transition to the slower mobility was found to have a characteristic time that is proportional to the electron lifetime in LXe, suggesting that impurities play a role in the process.
The fraction of $^{214}\mathrm{Bi}^{+}$ ions from $^{214}\mathrm{Pb}$ decay in LXe was also determined to be $f_\beta = 76.4 \pm 5.7\%$, assuming that the daughter ion fraction for the $^{218}\mathrm{Po}$ decay is the same as for the $^{222}\mathrm{Rn}$ decay. 
This result provides some basis for expecting high ionization fraction of $^{136}\mathrm{Ba}$ from double beta decay, which is relevant to the design of possible tagging methods.
Progress on some of these tagging methods have been reported recently \cite{Twelker2014,Mong2015}.
The higher daughter ion fraction than hole ion fraction, particularly for alpha decay, was discussed, and a potential mechanism of charge transfer reactions with Xe holes and/or Penning ionization reactions with excitons was proposed to explain this. 
The EXO-200 homogeneous LXe TPC has measured $e^-$ drift velocities of $1.7\times 10^6~\mathrm{mm/s}$ \cite{Albert2014_2vbb}, ion drift velocities of $\approx 1~\mathrm{mm/s}$ and neutral ion flows of $\approx10^{-2}~\mathrm{mm/s}$, spanning a velocity range of $10^8$.

\section{Acknowledgments}
EXO-200 is supported by DOE and NSF in the United States, NSERC in Canada, SNF in Switzerland, IBS in Korea, RFBR(14-22-03028) in Russia, CAS-IHEP Fund in China, and DFG Cluster of Excellence ``Universe'' in Germany. 
EXO-200 data analysis and simulation uses resources of the National Energy Research Scientific Computing Center (NERSC), which is supported by the Office of Science of the U.S. Department of Energy under Contract No. DE-AC02-05CH11231. 
The collaboration gratefully acknowledges the WIPP for their hospitality

\bibliography{alphaion_main}

%\appendix

\end{document}